\documentclass[aps, pre, twocolumn]{revtex4-2}

\usepackage{amsmath, amssymb}
\usepackage{graphicx}
\usepackage{xcolor}
\usepackage[normalem]{ulem}

\graphicspath{{./figures/}}

\begin{document}

\title{Prethermalization in Fermi-Pasta-Ulam-Tsingou chains}
\author{Gabriel M.~Lando}
\author{Sergej Flach}
\affiliation{Center for Theoretical Physics of Complex Systems, Institute for Basic Science (IBS), Daejeon, Korea, 34126}
\keywords{thermalization slowing-down, prethermalization, Lyapunov spectrum, ergodic hypothesis}

\begin{abstract}
The observation of the Fermi-Pasta-Ulam-Tsingou (FPUT) paradox, namely the lack of equipartition in the evolution of a normal mode in a nonlinear chain on unexpectedly long times, is arguably the most famous numerical experiment in the history of physics. Seventy years after the original publication, most studies in FPUT chains still focus on long wavelength initial states similar to the original paper. It is shown here that all characteristic features of the FPUT paradox are rendered even more striking if modes with short(er) wavelengths are evolved instead. Since not every normal mode leads to equipartition, we also provide a simple technique to predict which modes, and in what perturbation order, are excited starting from an initial mode (root) in $\alpha$-FPUT chains. The excitation sequences associated with a root are then numerically shown to spread energy at different speeds, leading to prethermalization regimes that become longer as a function of mode excitation number. This effect is visible in observables such as mode energies and spectral entropies and, surprisingly, also in the time evolution of invariant quantities such as Lyapunov times and Kolmogorov-Sinai entropies. Our findings generalize the original FPUT experiment, provide an original look at the paradox's source, and enrich the vast literature dedicated to studying equipartition in classical many-body systems.
\end{abstract}

\maketitle

\section{Introduction}

The numerical experiment of Fermi, Pasta, Ulam and Tsingou (FPUT) \cite{fermi1955studies} was simultaneously the first computational approach in the study of out-of-equilibrium many-body systems and a source of considerable headache shared by generations of physicists in the past 70 years \cite{fermi1955studies,ford1992fermi,fpugen,gallavotti2007fermi}. A proper understanding of FPUT's findings--which contradicted the standard knowledge of the time and were considered paradoxical--relies heavily on modern mathematical and physical tools unavailable to the original authors, e.g., a mature formulation of Kolmogorov-Arnol'd-Moser (KAM) theory \cite{kolmogorov1954conservation, arnold1963small, moser1962invariant}, resonance overlap criteria \cite{chirikov1979universal} and Birkhoff-Gustavson normalization \cite{birkhoff1927periodic, gustavson1966formal, ozorio1990hamiltonian, murdock2003normal, wiggins2003introduction}, and also on the power of modern computers. Sitting at the intersection of several fields, it is no surprise that this topic has drawn sustained interest from researchers in nonlinear dynamical systems and statistical mechanics.

The original FPUT experiment consists of evolving the fundamental mode of a harmonic chain in a weakly perturbed, nonintegrable system in which this mode is no longer a stationary state. More specifically, the authors studied the evolution of the first normal mode of a discretized string governed by the Hamiltonian
\begin{equation}\label{eq:FPUT}
    H(\boldsymbol{p},\boldsymbol{x}) = \sum_{j=0}^N \left[ \frac{p_j^2}{2} + \frac{(x_{j+1}-x_j)^2}{2} + \frac{\alpha (x_{j+1}-x_j)^3}{3} \right] \, ,
\end{equation}
where $p_j$ and $x_j$ are the momentum and position of each discretized mass in the chain, $\alpha{=}1/4$ and fixed boundary conditions were used, namely $x_0{=}x_{N+1}=0$. A very small energy density was chosen for the initial state, i.e., $h{=}H(\boldsymbol{p},\boldsymbol{x})/N \ll 1$, which limited the contribution of the cubic term and rendered the system a weak perturbation of the integrable harmonic chain obtained by setting $\alpha{=}0$. Even with such a weak perturbation, the dynamics under \eqref{eq:FPUT} is chaotic, which motivated the original expectation that the energy of the initially excited mode would rapidly spread among all other normal modes, leading to equipartition. However, what was observed was quite different: Not only did energy equipartition fail to occur over long timescales, but the system exhibited near recurrences, with energy periodically returning to the initial mode and many other modes remaining only weakly excited throughout the evolution \cite{pace2019behavior}.

Today, the convergence of numerous studies provides a coherent explanation for the originally observed lack of energy equipartition in finite, weakly anharmonic systems (for reviews, see \cite{ford1992fermi, poggi1997exact, berman2005fermi, campbell2005introduction, gallavotti2007fermi, benettin2008fermi}) -- although speculations on whether or not the original FPUT setup would eventually reach equilibrium have become a prominent topic in metaphysics. First, it is important to note that the energy density chosen by FPUT was extremely small; indeed, a subsequent investigation by Chirikov and Izrailev showed that larger energy densities do lead to energy equipartition \cite{chirikov1970statistical}. Second, KAM theory was later shown to apply to many FPUT-like systems, rigorously proving the existence of an energy threshold below which equipartition does not take place \cite{nishida1971note, rink2000near, rink2001symmetry, verhulst2020variations}. Third, and most relevant to this manuscript, is the fact that the initial condition used by FPUT is highly \emph{atypical}, corresponding to a stationary solution of the nearby harmonic chain \cite{casetti1997fermi}. Moreover, this atypical initial state lies close to exact, time-periodic stationary solutions of the anharmonic system, known as q-breathers \cite{flach2005q, flach2006q}.  This stands in sharp contrast to \emph{typical} initial conditions which, when expressed in the normal mode basis, involve combinations of all of its elements. Since these modes form the canonical action variables of the harmonic chain, perturbing a typical initial condition leads to an immediate coupling of all actions, whereas in the FPUT case, only a single action is perturbed. It is therefore natural that the resulting time evolution in the original FPUT study was highly atypical, reflecting the atypicality already present in the initial state itself.

A further important point is that atypical initial states are not all alike. Their time evolution depends sensitively on which initial mode, or \emph{root}, is excited. The dynamics of roots is characterized by excitation sequences that spread in mode space and, depending on relatively simple algebraic rules, completely or only partially fill it. The case of complete filling corresponds to \emph{thermal} roots, for which time-dependent observables generally reach equipartition. In contrast, if the root excites only a subset of modes, known as its \emph{bush} \cite{chechin2002bushes}, then equipartition fails to occur, and the corresponding root is termed \emph{nonthermal}.

A particularly striking example of nonthermal roots is provided by q-breathers: families of exact, time-periodic solutions that remain localized in mode space and reduce to a single-mode excitation as the anharmonicity tends to zero \cite{flach2005q, flach2006q}. These states show that certain normal modes and their perturbations can remain stationary even in the presence of nonlinearity, offering remarkable counterexamples to the expectation that observables in nonlinear classical many-body systems necessarily undergo some form of equilibration. They also reveal that ergodicity-breaking behavior can arise not only as a function of decreasing energy density, as in the original FPUT experiment, but also as a consequence of the choice of initial state, as we shall explore in detail here. 

This manuscript is devoted to investigating the properties of atypical initial conditions (normal modes) in FPUT chains, and also comparing them to the ones of typical (random) ones. We provide explicit algebraic rules to determine which modes are excited given an initial root, together with their short-time perturbation order in $\alpha$. We then numerically investigate the time evolution of thermal roots of several different excitation numbers, showing that the paradox associated to the fundamental mode is substantially maximized for higher excited roots, e.g., the prethermal regime and its metastable plateaus become increasingly longer as a function of root excitation number for a fixed energy density. This is shown both by a direct tracking of mode energies as a function of time and by computing equipartition times from spectral entropies.

We also show that the root excitation number impacts quantities that do not depend on the initial state. Indeed, albeit Lyapunov times and Kolmogorov-Sinai (KS) entropies being invariant with respect to a large number of transformations and being constant in an ergodic system regardless of the initial state used for computing them \cite{eichhorn2001transformation}, we show here that their \emph{transient} properties are strikingly different from those of typical initial states and clearly reflect the fact that the system is trapped in phase space \cite{casetti1997fermi}. More specifically, both the maximal Lyapunov exponent (the inverse of the Lyapunov time) and the Kolmogorov-Sinai entropy show a dip in their time evolution when computed from roots, which is absent if the initial state chosen is typical. Moreover, the duration and depth of this dip increases with excitation number, showing that excited modes remain trapped for even longer in near-integrable portions of phase space, likely due to their proximity to q-breathers.

The paper is structured as follows. Sec.~\ref{sec:model} discusses the model, which is entirely focused in \eqref{eq:FPUT}, introducing normal modes, spectral entropies and Lyapunov data which will be thoroughly employed in the rest of the paper. Sec.~\ref{sec:sequences} presents the rules for computing excitation sequences for roots, which are then evolved in an FPUT chain with $N{=}63$ masses in Sec.~\ref{sec:numerics}. A discussion of our findings can be found in Sec.~\ref{sec:discussion}, together with a conclusion in Sec.~\ref{sec:conclusion}. An appendix is also included providing examples of pathological numerical behavior due to bush instability.

\section{Model and observables}\label{sec:model}

We restrict our analysis to cubic FPUT chains, also known as the $\alpha$-FPUT model, with Hamiltonian \eqref{eq:FPUT}. In the following we recall how to rewrite the FPUT Hamiltonian in normal mode (or phonon, or eigenmode) basis, and introduce quantities such as mode energies, spectral entropies and equipartition times. We then describe how to obtain important invariant quantities that will be used in the following sections, such as Lyapunov times and Kolmogorov-Sinai entropies.

\subsection{Normal mode basis}

Since Eq.~\eqref{eq:FPUT} has $N$ degrees of freedom its harmonic limit has a set of $N$ normal modes, denoted $\{P_j,Q_j\}_j$, with which we can rewrite the position and momentum in Eq.~\eqref{eq:FPUT} at any time $t$ as
\begin{align}\label{eq:mode_coordinates}
    \begin{pmatrix}
        p_j(t) \\
        x_j(t)
    \end{pmatrix}
    = \sqrt{\frac{2}{N+1}} \sum_{n=1}^N 
    \begin{pmatrix}
        P_j(t) \\
        Q_j(t)
    \end{pmatrix}
    \sin \left( \frac{\pi j n}{N+1}\right) \, ,
\end{align}
each mode having frequency and energy
\begin{align}\label{eq:freqs_and_energies}
    \omega_j{=}2 \sin \left[ \frac{\pi j}{2(N+1)} \right] \, , \quad E_j = \frac{P_j^2 + \omega_j^2 Q_j^2}{2} \quad . 
\end{align}
In these canonical coordinates, the Hamiltonian \eqref{eq:FPUT} takes the form \cite{nishida1971note}
\begin{align}\label{eq:FPUT2}
    &H(\boldsymbol{P},\boldsymbol{Q}) = \frac{1}{2} \sum_{j=1}^N \big( P_j^2 + \omega_j^2 Q_j^2 \big) \notag \\
    &\quad + \frac{\alpha}{3\sqrt{2(N+1)}} \sum_{i,j,k=1}^N  B_{i,j,k} \, \omega_i \, \omega_j \, \omega_k \, Q_i Q_j Q_k \, ,
\end{align}
where \cite{flach2008fermi}
\begin{equation}\label{eq:coupling_coefficient}
    B_{i,j,k} = \delta_{i+j,k} + \delta_{k+i,j} + \delta_{j+k,i} - \delta_{i+j+k,2(N+1)} \, .
\end{equation}
Note that $B_{i,j,k}$ is invariant with respect to permutations of its indices. 

\subsection{Spectral entropy}\label{subsec:spectral-entropy}

Equation \eqref{eq:FPUT} is much easier to numerically deal with than \eqref{eq:FPUT2}. However, we are interested in several properties that need to be computed in the normal mode basis, such as the mode energies themselves and the [normalized] spectral entropy 
\begin{equation}\label{eq:spectral_entropy}
    \eta(t) = \frac{S(t) - \log N}{S(0) - \log N} \, ,    
\end{equation}
where the Shannon entropy and normalized energies are given, respectively, by
\begin{equation}\label{eq:entropies_and_energies}
    S(t) = - \sum_{j=1}^N \rho_j(t) \log \rho_j(t) \, , \quad \rho_j(t) =  \frac{E_j}{\sum_{j=1}^N E_j } \, .
\end{equation}

Spectral entropy has been used several times to quantify equipartition in nonlinear systems \cite{casetti1997fermi, gallavotti2007fermi, cretegny1998localization, danieli2017intermittent}. The logic for employing it is simple: If equipartition is reached, then $\rho_j(t)$ will be stationary and the same for every mode $j$, freezing $\eta(t)$ at some value $\langle \eta \rangle$. If one assumes a Gibbs distribution at equilibrium, then $\langle \eta \rangle$ can be calculated analytically as a function of the Shannon entropy of the initial state and is given by
\begin{equation}\label{eq:equilibrium_eta}
    \langle \eta \rangle = \frac{1 - \gamma}{\log N - S(0)} \, ,  
\end{equation}
where $\gamma{\approx}0.5772$ is the Euler constant \cite{goedde1992chaos}. Since $\eta(0){=}1$ for any initial state, one expects $\eta(t)$ to decrease and follow a complicated and possibly highly oscillatory evolution until eventually hitting its stationary value. Once this happens the dynamics cannot, at least on average, depart significantly from thermal oscillations around $\langle \eta \rangle$. It is then natural to consider that equipartition is reached once $\eta(t)$ hits its equilibrium value for the first time \cite{danieli2017intermittent}. Such first-passage time furnishes our definition of \emph{equipartition time}, $\tau_\mathrm{eq}$, and is appealing because it does not require any \emph{ad hoc} assumptions. 

\subsection{Lyapunov invariants and Kolmogorov-Sinai entropy}

Equipartition time as defined in Sec.~\ref{subsec:spectral-entropy} is, evidently, attached to the observables being monitored, namely the mode energies. Different observables will generally reach equipartition at different times (if at all), such that observable equipartition does not actually measure properties intrinsic to the system. Nevertheless, since mode energies are conserved in the unperturbed harmonic chain, one knows well how these observables should behave upon approaching an integrable limit, i.e., equipartition should slow down and eventually stop for sufficiently small energy densities. 

There is, however, an observable-independent timescale that is completely intrinsic to the system and should not depend on the initial state, given by the Lyapunov time (the inverse of the maximal Lyapunov exponent). Since, like the maximal Lyapunov exponent (MLE), all exponents in the Lyapunov spectrum (LS) are invariant with respect to homeomorphisms \cite{eichhorn2001transformation}, the LS provides a valuable set of timescales related to the average ``pull'' along each 1-dimensional submanifold that forms the hyperbolic skeleton of the system's dynamics in phase space \cite{oseledec1968multiplicative}. The information in the LS can also be condensed in the form of a scalar, the Kolmogorov-Sinai entropy, which by Pesin's theorem can be obtained as the sum of all positive exponents in the LS \cite{pesin1977characteristic}.

To compute the LS we employ the well-known prescription of \cite{benettin1980lyapunov, geist1990comparison}, which amounts to computing the spectrum of the symmetric matrix
\begin{equation}\label{eq:stability-matrix}
    \Gamma(\boldsymbol{p}_0,\boldsymbol{x}_0)= \lim_{t \to \infty} \left[ \mathcal{M}^T(\boldsymbol{p}_0,\boldsymbol{x}_0;t) \mathcal{M}(\boldsymbol{p}_0,\boldsymbol{x}_0;t) \right]^{1/2t} \, .
\end{equation}
In the above, $\mathcal{M}$ represents the system's stability matrix, obtained by solving Hamilton's equations in tangent space:
\begin{equation}
    \frac{\mathrm{d}\mathcal{M}(\boldsymbol{p},\boldsymbol{x};t)}{\mathrm{d}t} = \mathcal{J} \mathrm{Hess} \left[ H(\boldsymbol{p},\boldsymbol{x}) \right] \mathcal{M}(\boldsymbol{p},\boldsymbol{x};t) \, ,
\end{equation}
where $\mathrm{Hess}$ is the Hessian with respect to $(\boldsymbol{p},\boldsymbol{x})$. Since the stability matrix has to be computed along a trajectory, the above equation is solved in parallel with Hamilton's equations for an initial phase-space point $(\boldsymbol{p}_0,\boldsymbol{x}_0)$ until a final time long enough to result in a converged LS. During this evolution, one must also perform QR-diagonalizations of $\mathcal{M}$ to avoid numerical denegenacies, as is well-known and described in, e.g., Ref.~\cite{geist1990comparison}.

As can be seen in \eqref{eq:stability-matrix}, the stability matrix will generally depend on the trajectory along which it is calculated, i.e., on the initial phase-space point $(\boldsymbol{p}_0, \boldsymbol{x}_0)$. However, for a sufficiently large number of degrees of freedom and energy density, phase space will consist of a single metrically intransitive set that is equally accessible to all trajectories in the system \cite{abraham2008foundations, palis2012geometric}. Equivalently, one can formulate this previous property as stating that, for sufficiently large $N$ and $h{=}H(\boldsymbol{p}, \boldsymbol{x})/N$, the system lies outside the KAM regime in which phase space can be split into a chaotic web and pockets of near-integrability. Thus, all trajectories are ``free to roam'' around the entire accessible phase space, and the LS is expected to be independent of the trajectory along which it is computed. This independence is a stronger evidence of ergodicity than any conclusion drawn from time-dependent observables, and can be verified more succinctly by comparing KS entropies instead of full LS.

\section{Excitation sequences}\label{sec:sequences}

We start this section by writing an abridged version of Newton's second law for a given mode $i$, obtained from \eqref{eq:FPUT2}:
\begin{align}\label{eq:newtons_law}
    \ddot{Q}_i + \omega_i^2 Q_i \sim \alpha \sum_{j,k=1}^N B_{i,j,k} \, Q_j Q_k \, .
\end{align}
Thus, the mode $Q_i$ is excited by a single mode $Q_j$ or a mixture of modes $Q_j$ and $Q_k$. Excitations from single modes have $j{=}k$ and the coupling tensor assumes the form
\begin{align}\label{eq:lone}
    B_{i,j,j} &= \delta_{i+j,j} + \delta_{j+i,j} + \delta_{2j,i} - \delta_{i+2j,2(N+1)} \notag \\
    \Longrightarrow \quad B_{i,j,j} &= \delta_{i,2j} - \delta_{i,2(N+1)-2j} \, ,
\end{align}
where the first two terms vanished because $j$ and $i$ are never zero. Now, for mixed excitations we have $j{\neq}k$, so
\begin{align}\label{eq:mixed}
    B_{i,j,k} &= \delta_{i+j,k} + \delta_{k+i,j} + \delta_{j+k,i} - \delta_{i+j+k,2(N+1)} \notag \\
    \Longrightarrow \quad B_{i,j,k} &= \delta_{i,k+j} + \delta_{i,|k-j|} - \delta_{i,2(N+1)-(j+k)} \, .
\end{align}
The above expression is composed of an ascending term, namely $\delta_{i,k+j}$, which dictates which higher-order mode to excite starting from $j$ and $k$; a right reflection term, $\delta_{i,2(N{+}1){-}(j{+}k)}$, which dictates which term to excite if $j{+}k{>}(N{+}1)$, i.e., if the next mode is too large and leaves mode space; and a descending term, $\delta_{i,|k-j|}$, which propagates the excitation backward in mode space and is also responsible for controlling left reflections, which take place when the next excited mode is smaller than zero. 

\subsection{First excited mode and q-breathers}

Initializing the system in a root $Q_r$ means setting $Q_j{=}0$ unless $j{=}r$. The first excitation cycle, therefore, starts with a lone mode $Q_r$, which by \eqref{eq:lone} excites the modes $i{=}2r$ and $i{=}2(N{-}r{+}1)$. If the first condition is met, then the second delta will be one iff $r{=}(N{+}1)/2$, with $N{+}1$ even. Thus, if the chain has an odd number of sites and the chosen root is $Q_{(N+1)/2}$, then all coefficients of the coupling tensor vanish and the energy remains forever localized in the root: this mode is a q-breather.

Now, assuming that only the second Kronecker delta in \eqref{eq:lone} vanishes, then the only mode that is excited starting from $Q_r$ is $Q_{2r}$ \cite{bivins1973nonlinear, sholl1990modal, flach2006q}. If, however, only the first delta vanishes, then the first excited mode is obtained after a reflection and is given by $Q_{2(N-r+1)}$. These scenarios are fundamentally identical but, for simplifying writing excitation sequences, we will always assume that the root mode fulfills $2(r{-}1){<}N$, so that the first excited mode will always be $Q_{2r}$.

Moving foward, a na{\"i}ve perturbative argument allows us to estimate how the amplitude of $Q_{2r}$ scales with the anharmonicity parameter, $\alpha$, for short times. It goes as follows: Since $Q_r$ carries all the initial energy in the system and $\alpha$ is small, the solution $Q_r(t)$ starts close to a sine or a cosine, as can be seen by setting $\alpha{=}0$ in \eqref{eq:newtons_law}. This approximation scales as $\mathcal{O}(\alpha^0)$ \cite{flach2006q}. Now, $Q_{2r}$ emerges from the product $Q_r Q_r$ in \eqref{eq:newtons_law}, which is also of 0th order in $\alpha$, but the $\alpha$ multiplying the right-hand side of \eqref{eq:newtons_law} gives $Q_{2r}(t)$ a $\mathcal{O}(\alpha^1)$ dependence for short times \cite{sholl1991recurrence}.

\subsection{First ascending sequence}

The second excited mode is obtained from $Q_r$ and $Q_{2r}$. Evidently, $Q_{2r}$ excites $Q_{4r}$ due to \eqref{eq:lone} and, since $Q_{2r}$ is of $\mathcal{O}(\alpha^1)$, the amplitude of $Q_{4r}(t)$ goes as $\alpha \, Q_{2r} Q_{2r}=\mathcal{O}(\alpha^4)$ for short times. Mode $Q_{3r{=}r+2r}$ is also excited according to \eqref{eq:mixed}, but now the short-time amplitude will scale as $\alpha \, Q_{r} Q_{2r}=\mathcal{O}(\alpha^2)$. By induction, the first ascending sequence for a root $Q_r$ is given by  
\begin{align}
    &\{ Q_r , Q_{2r} , Q_{3r}, Q_{4r}, Q_{5r}, \dots, Q_{a r}\} \, .
\end{align}
It includes all multiples of the root mode ordered by how their short-time dynamics scales with $\alpha$, namely $\mathcal{O}(\alpha^{a-1})$, until a reflection takes place in \eqref{eq:mixed}, i.e., until $j{+}k {=} 2(N{+}1)$ for some $Q_j$ and $Q_k$. The reflection, however, will only excite new modes if $2(N{+}1){-}(j{+}k)$ is not a multiple of $r$, since otherwise the terms would have already been excited during the ascending sequence. 

\subsection{First descending sequence}\label{subsec:first_descending_sequence}

At the end of the first ascending sequence every term is a multiple of the root, $r$. Thus, at the reflection, substituting $j{+}k{=}a \, r$ in the last delta of \eqref{eq:mixed} singles out $r{=}2(N{+}1)/a$, allowing us to state two important facts:
\begin{enumerate}
    \item If $r$ divides $2(N{+}1)$, then the first ascending sequence is reflected into itself. This shows, in particular, that $Q_2$ is nonthermal for all $N$, since $2(N{+}1)$ is a always divisible by 2 \cite{bivins1973nonlinear, sholl1990modal}. It also shows that $Q_1$ is a \emph{trivial} thermal root, since it excites all modes in the first ascending sequence for any $N$;
    \item The reflection takes place when a multiple of $r$ is larger than $N$ for the first time, i.e., when $Q_j=Q_{(a+1)r}$. Evidently this mode lies outside mode space and the excited mode is actually $Q_{2(N+1)-(a+1)\,r}$ after the reflection. A consequence is that even roots can never excite odd modes, since for even modes $2(N{+}1){-}(a{+}1)\,r$ is always even \cite{sholl1990modal}. This shows that $Q_{2r}$ are nonthermal roots for all $r$ and all $N$.
\end{enumerate}
Once a reflection occurs, the selection rule for $|j{-}k|$ generates a descending sequence starting from $Q_{2(N+1)-(a+1)\,r}$, namely,
\begin{equation}
    (Q_{2(N+1)-(a+2)\,r}, Q_{2(N+1)-(a+3)\,r}, \dots ) \, .
\end{equation}
The descent continues in multiples of $r$ until the next mode would exit the chain, i.e., until $2(N+1)-(a+b)\,r < 0$. Despite the reflection, it is important to note that each term in the descending sequence comes from a product like $Q_{a r} Q_{b r}$, and will therefore have an amplitude that scales perturbatively as $\mathcal{O}(\alpha^{a+b-1})$ for short times, just as in the ascending sequence. This shows both ascending and descending sequences can be tied together and monotonically ordered based on how their short-time solutions scale with $\alpha$.

\subsection{Full excitation sequences}

Once the first descending sequence is over, another reflection takes place, now going from left to right, mediated by $\delta_{i,|j-k|}$ in \eqref{eq:mixed}. Substituting the last mode number from the descending sequence in this delta, we see that the excited mode after this reflection is $Q_{2(N+1)-(a+b)\,r + r}$. The second ascending sequence will start from this mode number, once again ascending in steps of $r$ until $2(N+1)-(a+b-c)r > N$. Then, we have another reflection from left to right that excites mode $2(N+1) - [2(N+1) - (a+b-c)r-r]=(a+b-c+1)r$, and so on. Instead of writing full abstract sequences, we now provide a few examples showing how easy it is to write them in practice. 

\begin{figure}
    \centering
    \includegraphics[width=\linewidth]{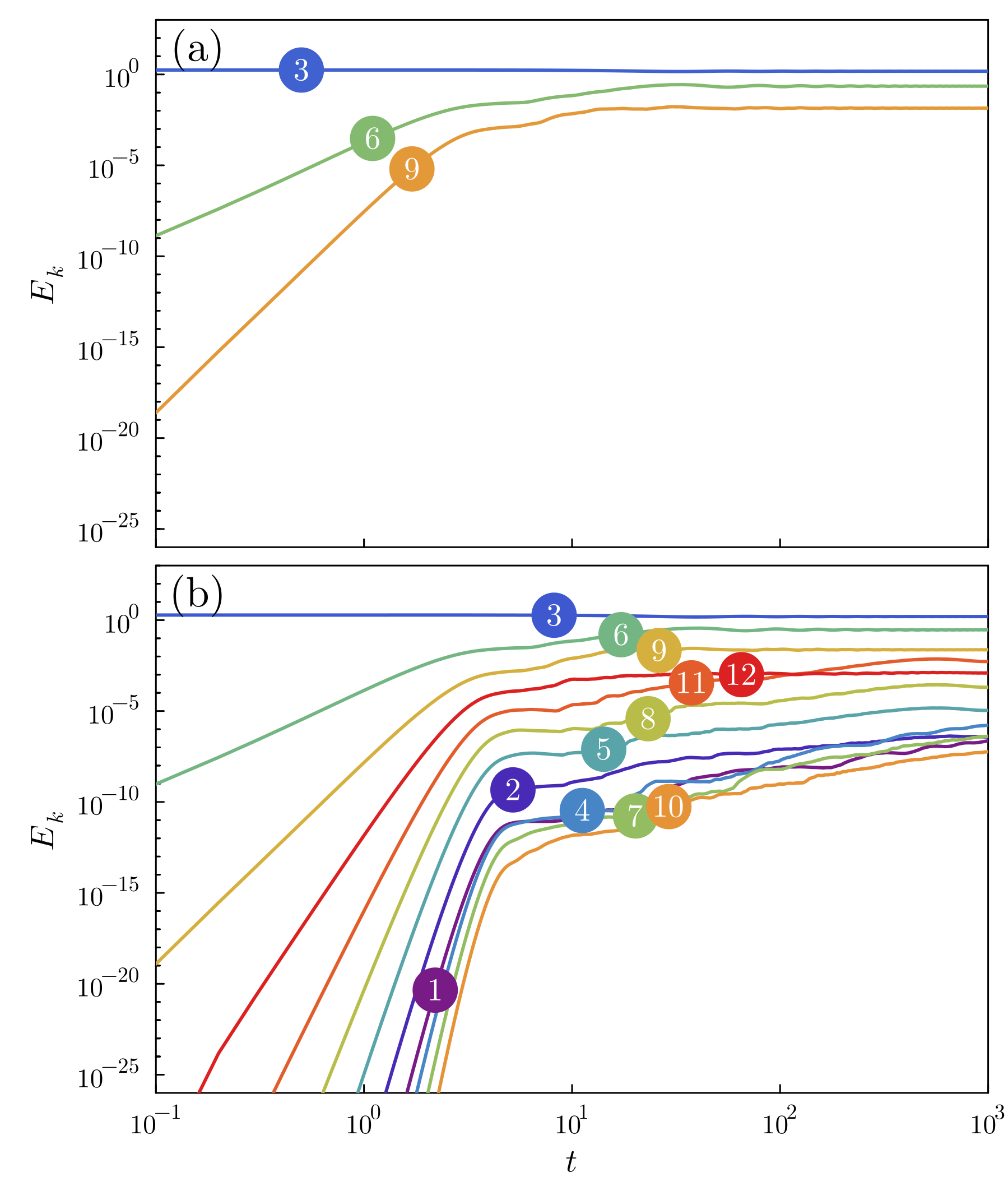}
    \caption{Evolution of mode energies for the root $Q_3$ in an $\alpha$-FPUT chain with (a) $N{=}11$ and (b) $N{=}12$ masses, both for $h{\approx}0.16$ and $\alpha{=}1/4$. The root $Q_3$ is only thermal in the latter case, while in the former its bush is composed of all multiples of $3$ that are smaller than $11$. The mode excitation sequence in (b) is the same as the one predicted using the coupling coefficients, \eqref{eq:mode_sequence2}.}
    \label{fig:sequences}
\end{figure}

Let us start with a chain with $N{=}11$ masses. Since $2(N{+}1){=}24$, the only three nontrivial thermal roots in this chain are $Q_5$, $Q_7$ and $Q_{11}$. The root $Q_3$ is evidently nonthermal since it divides 24 (c.f.~Sec.~\ref{subsec:first_descending_sequence}), but let's take a look at its excitation sequence. Denoting a sequence starting from root $r$ in a system with $N$ normal modes by $S_r^N$, the first ascending sequence is $S_3^{11}=\{Q_3, Q_6, Q_{9} \}$, and we stop here because the next term falls outside mode space. The reflection term comes from $i{=}2(N{+}1){-}(6{+}9)=9$, which is already excited. This will be now propagated backwards to $Q_6$ and $Q_3$ accoding to \eqref{eq:mixed}, retracing the ascending sequence and never exciting any new mode. Now, for the leftmost reflection, the excited mode is $Q_{|3-3|}$, which does not exist. Thus, the bush of $Q_3$ is given by $\{Q_3, Q_6, Q_9 \}$. This contradicts a statement made in \cite{sholl1990modal}, which states that all odd modes should be excited if $r$ and $N$ are relatively prime, but is verified numerically in Fig.~\ref{fig:sequences}(a).

Let us now consider what happens to $Q_3$ for $N{=}12$ instead. The first ascending sequence is now updated to $\{Q_3, Q_6, Q_{9}, Q_{12} \}$. The reflection gives $i{=}2(N{+}1){-}(12{+}3){=}11$, which was not previously excited and is not a multiple of $r$. The next excited modes will then be $11{-}3{=}8$, $11{-}6{=}5$, etc, and the full excitation sequence is
\begin{align}\label{eq:mode_sequence1}
    S_3^{12} &= \{Q_3, Q_6, Q_{9}, Q_{12}, Q_{11=26-(12+3)}, \notag \\
    &\qquad  Q_{8}, Q_{5}, Q_{2}, Q_{1=|2-3|}, Q_{4}, Q_{7}, Q_{10} \} \, ,
\end{align}
where modes are ordered by how their short-time solutions scale with $\alpha$. This excitation sequence is seen to match numerics in Fig.~\ref{fig:sequences}(b). 

As another example, for $S_3^{13}$ we have
\begin{align}\label{eq:mode_sequence2}
    S_3^{13} &= \{Q_3, Q_6, Q_{9}, Q_{12}, Q_{13=28-(12+3)}, \notag \\
    &\qquad Q_{10}, Q_{7}, Q_{4}, Q_{1}, Q_{2=|1-3|}, Q_{5}, Q_{8}, Q_{11} \} \, ,
\end{align}
and, for $S_5^{13}$, 
\begin{align}\label{eq:mode_sequence3}
    S_3^{13} &= \{Q_5, Q_{10}, Q_{13=28-(10+5)}, Q_{8}, Q_{3}, Q_{2=|3-5|},  \notag \\
    &\qquad  Q_{7}, Q_{12}, Q_{11=28-(12+5)}, \notag \\ 
    &\qquad Q_{6}, Q_{1}, Q_{4=|1-5|}, Q_{9} \} \, .
\end{align}

\subsection{Mode connections and amplitudes}\label{subsec:graphs}

Excitation sequences such as \eqref{eq:mode_sequence2} and \eqref{eq:mode_sequence3}, and also the trivial sequence generated by $Q_1$, completely fill mode space for $N{=}13$. However, these sequences do not provide insight into how energy is exchanged between excited modes -- an aspect we will briefly explore in the following.

Starting from \eqref{eq:mode_sequence2}, we see that the pair that most strongly connects to $Q_{13}$ in $S_5^{13}$ is $(Q_5, Q_{10})$, since the right-hand side of \eqref{eq:newtons_law} gives $\alpha \, Q_5 Q_{10}=\mathcal{O}(\alpha^2)$. There are also connections formed by other pairs, such as $Q_5$ and $Q_8$, but these connections are of higher order in $\alpha$, e.g., $\alpha \, Q_8 Q_5 = \mathcal{O}(\alpha^4)$. Thus, the short-time regime will see $Q_{13}$ exchanging energy primarily with $Q_5$ and $Q_{10}$. The situation is similar in $S_3^{13}$ in \eqref{eq:mode_sequence3}, where $(Q_{12},Q_3)$ and $(Q_9,Q_6)$ both connect to $Q_{13}$ and have weighs of $\mathcal{O}(\alpha^4)$. The case of $S_1^{13}$ is rather extreme, since the sum or difference of any two mode numbers that equals 13 or 15 all connect to $Q_{13}$, and all connections scale as $\mathcal{O}(\alpha^{12})$. 

The argument above sheds light on why energy remains localized in a subset of modes for short times, but the simplified form of \eqref{eq:newtons_law} glosses over the multiplier $\omega_i \omega_j \omega_k$ in \eqref{eq:FPUT2}. These frequencies $\omega$, as seen in \eqref{eq:freqs_and_energies}, increase monotonically for higher mode numbers. A consequence is that, despite $Q_3$ in $S_1^{13}$ and $Q_{13}$ in $S_5^{13}$ both scaling perturbatively as $\mathcal{O}(\alpha^2)$, the amplitude of $Q_{13}$ in $S_5^{13}$ will lie above the one of $Q_3$ in $S_1^{13}$ for short times if the roots are excited with the same energy density. The same conclusion could be drawn from noticing that, for a fixed energy density, energy has to be shared among a smaller subset of excited modes in $S_5^{13}$ when compared to $S_1^{13}$ in the short-time regime, rendering local energies larger. These larger energies take longer to spread to other modes, since $Q_{13}$ in $S_5^{13}$ has exactly the same \emph{number} of connections as $Q_3$ in $S_1^{13}$ (and they are also of the same order). Note also that the next thermal root in $S_r^{13}$ would be $r{=}7$, but this mode is a q-breather for $N{=}13$. Thus, increasing mode number is tantamount to approaching a state for which energy does not spread in mode space, which offers a more qualitative way of interpreting the slowing down of energy equipartition for increasing $r$. 

\section{Numerical experiments}\label{sec:numerics}

\begin{figure*}
    \centering
    \includegraphics[width=\linewidth]{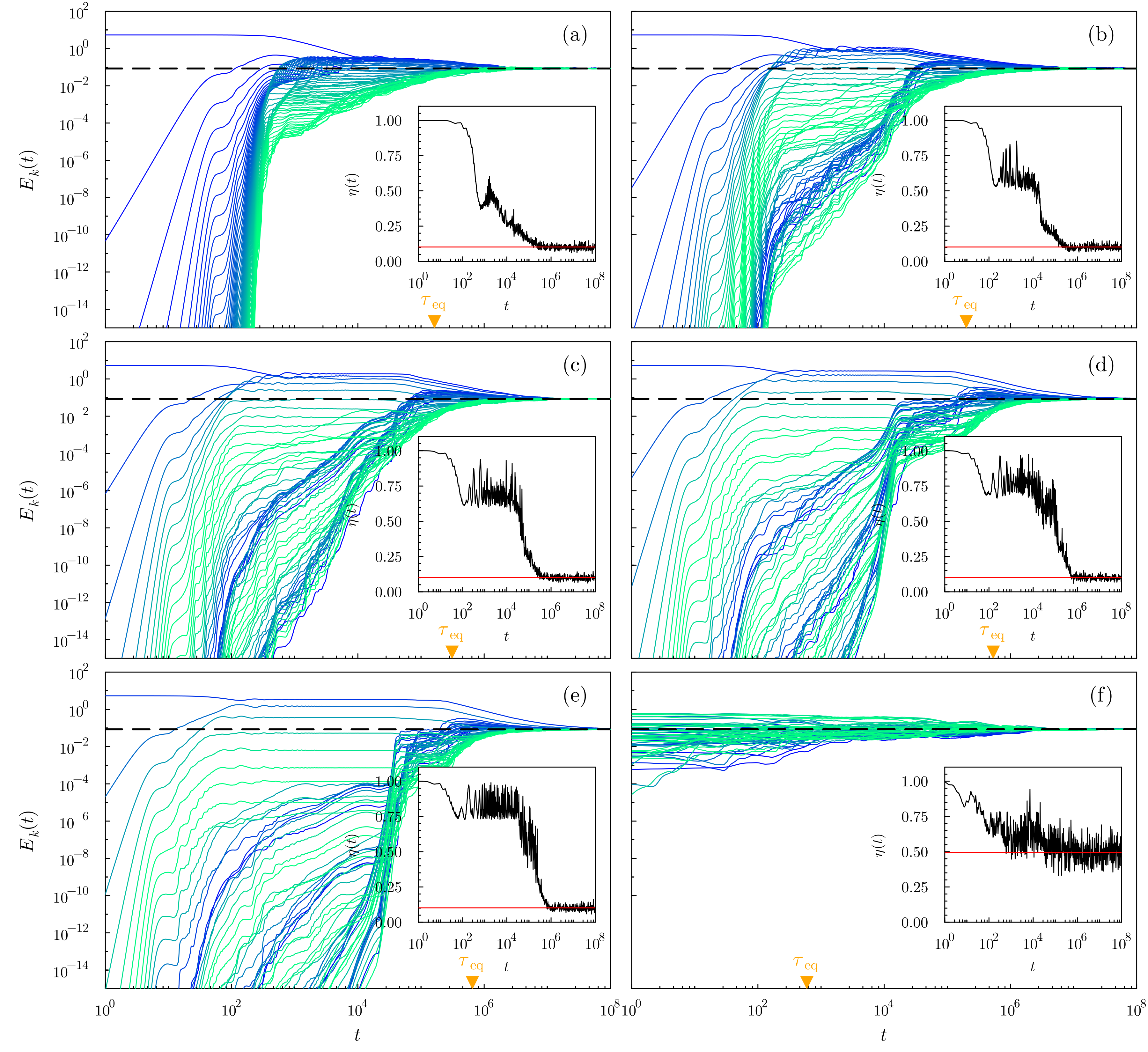}
    \caption{Evolution of mode energies for the first five odd modes, namely: (a) $Q_1$; (b) $Q_3$; (c) $Q_5$; (d) $Q_7$; (e) $Q_9$; and (f) a random initial condition, all with energy density $h{=}H(\boldsymbol{p}, \boldsymbol{x})/N{\approx}0.085$} in an $\alpha$-FPUT chain with $N{=}63$ and $\alpha{=}1/4$. Line colors correspond to mode frequencies, going from lower (blue) to higher (green), with their equilibrium value shown as a black dashed line. Equipartition times (orange triangles) are defined as the instant the time-evolving spectral entropy, shown in black in the inset, first attains its equilibrium value $\langle \eta \rangle$, given by the horizontal red line.
    \label{fig:mode_dynamics}
\end{figure*}

\begin{figure*}
    \centering
    \includegraphics[width=\linewidth]{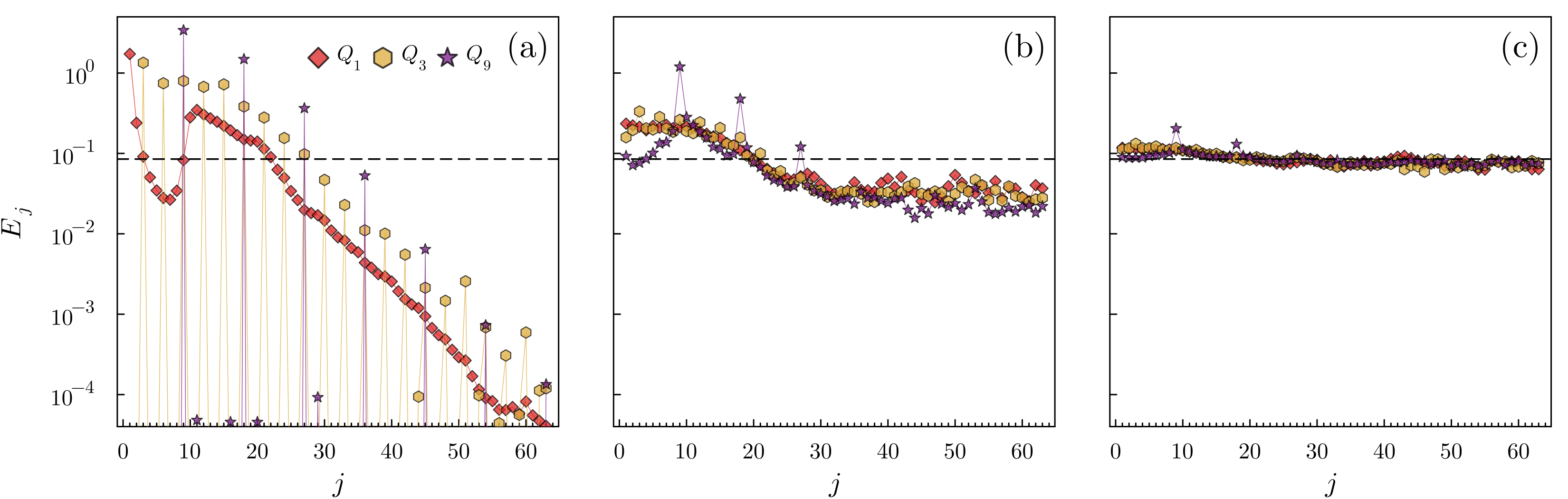}
    \caption{Mode energies, $E_j$, as a function of mode number, $j$, for roots $Q_1$, $Q_3$ and $Q_9$ for fixed times, with their equilibrium (average) value, $h$, shown as a black dashed line. Each panel is a vertical slice of a panel in Fig.~\ref{fig:mode_dynamics} at (a) $t{=}\tau_\mathrm{eq}/100$; (b) $t{=}\tau_\mathrm{eq}$ and (c) $t{=}10\tau_\mathrm{eq}$ (note that $\tau_\mathrm{eq}$ is different for each root). The short time of (a) allows for a clear visualization of the energy cascade flowing between modes that are multiples of 1, 3 and 9, as described in Sec.~\ref{sec:sequences}. At equipartition most of the site energies have the same order of magnitude, as seen in (b), with stronger mixing still taking place at high mode numbers while the lower half of the excitation sequence remains slightly isolated from the other modes. Even after ten equipartition times the nearly homogeneous energy distribution hints at which root was initially excited, as can be seen in the above-average energies of modes $Q_9$ and $Q_{18}$ for the root $Q_9$ in (c).}
    \label{fig:mode_energies_per_site}
\end{figure*}

In this section we perform computations in the $\alpha$-FPUT chain \eqref{eq:FPUT} with $N{=}63$ sites. This value of $N$ is chosen because $2(N{+}1){=}128$ is not divisible by any odd integer other than 1, such that all odd roots are thermal for this chain. Although initial states, final propagation times and energy densities will vary, evolution is always computed using an optimized second order symplectic integrator with step size $\Delta t{=}0.2$ \cite{mclachlan1992accuracy}. This results in a maximum relative energy error of around $10^{-3}$ for the largest energy densities chosen, and much smaller for most of them. We will focus on the evolution of random initial conditions and the first five odd roots $Q_1$, $Q_3$, $Q_5$, $Q_7$ and $Q_9$. The sampling of all initial conditions starts by setting $\boldsymbol{x}(0){=}0$. Then, for random states, each momentum component is drawn from a Maxwell-Boltzmann distribution, while for mode $Q_k$ we have momentum components fixed as $p_j(0){=}\sin[\pi k j/(N{+}1)]$. Once sampled, momenta are then renormalized in order to achieve the desired energy density for both types of initial conditions.

\subsection{Metastable plateaus}\label{subsec:plateaus}

Sec.~\ref{sec:sequences} described how energy can remain localized within a small subset of modes before spreading throughout mode space: the higher excited the root, the smaller the subset of excited modes and the larger their amplitudes, such that more time is needed in order to distribute their energy among other modes. During this transfer time the dynamics is mostly localized within the subset of excited modes, which essentially ``wait'' and form the metastable plateaus prominent in FPUT literature \cite{benettin2008fermi, matsuyama2015multistage, pace2019behavior}. 

The well-known fact that the metastable plateaus associated to roots become more visible at lower energy densities is unsurprising, since in the limit of zero energy density such modes are normal modes of the associated harmonic lattice. However, the behavior of plateaus for excited roots was not previously addressed. To this end, Fig.~\ref{fig:mode_dynamics} shows a comparison of evolving mode energies and spectral entropies for the $\alpha$-FPUT chain and initial states described in the introduction to this section. Fig.~\ref{fig:mode_dynamics}(a) displays the standard mode energy dynamics found in literature: energy initially placed in $Q_1$ monotonically spreads to all other modes and eventually results in energy equipartition, with all $E_k$'s converging to the mean energy $h=H(\boldsymbol{p},\boldsymbol{x})/N$ (black dashed line). Metastable plateaus are clearly visible in the prethermal regime, in which a measurement would indicate that the system is not ergodic and cannot be described by equilibrium statistical mechanics. This statement is evidently incorrect, as waiting longer would result in a system that does display statistical behavior. 

The approach from an out-of-equilibrium initial state towards a thermal one is most clearly seen in tracking the spectral entropy as a function of time, shown in the insets: the moment it touches the red line, which represents its equilibrium value \eqref{eq:equilibrium_eta}, is where we consider equipartition to have been achieved. The associated equipartition times are displayed in the main plots as an orange triangle and slowly increase as we move from $Q_1$ up to $Q_9$, with the shortest time corresponding to the random initial condition. Evidently, this latter case is less interesting since the equipartition time depends strongly on the initial sampling, but for roots the structures formed by mode energies as they evolve in time show that, indeed, higher-excited roots isolate an increasingly smaller subset of modes in mode space. Fig.~\ref{fig:mode_dynamics}(a) also shows a dense subset of modes homogeneously concentrated around the mean, while Fig.~\ref{fig:mode_dynamics}(e) clearly shows that two modes, namely $Q_9$ and $Q_{18}$, remain relatively isolated from the others for much longer times. To see this more clearly, in Fig.~\ref{fig:mode_energies_per_site} we take vertical slices of Fig.~\ref{fig:mode_dynamics} at three different fixed times for $Q_1$, $Q_3$ and $Q_9$. At $t{=}\tau_\mathrm{eq}/100$ the contents of Sec.~\ref{subsec:graphs} can be clearly visualized, namely the fact that increasing root number traps energy in a smaller subset of modes with higher amplitudes in the short-time regime. At $t{=}\tau_\mathrm{eq}$ low frequency modes that are multiples of 3 and 9 can still be seen to have above-average energies for roots $Q_3$ and $Q_9$. At $t{=}10\tau_\mathrm{eq}$ this effect has already dissipated for $Q_3$, but the energies of modes $Q_9$ and $Q_{18}$ remain slightly above the mean energy when evolving $Q_9$, as visible in Fig.~\ref{fig:mode_energies_per_site}(c).

\begin{figure}
    \centering
    \includegraphics[width=\linewidth]{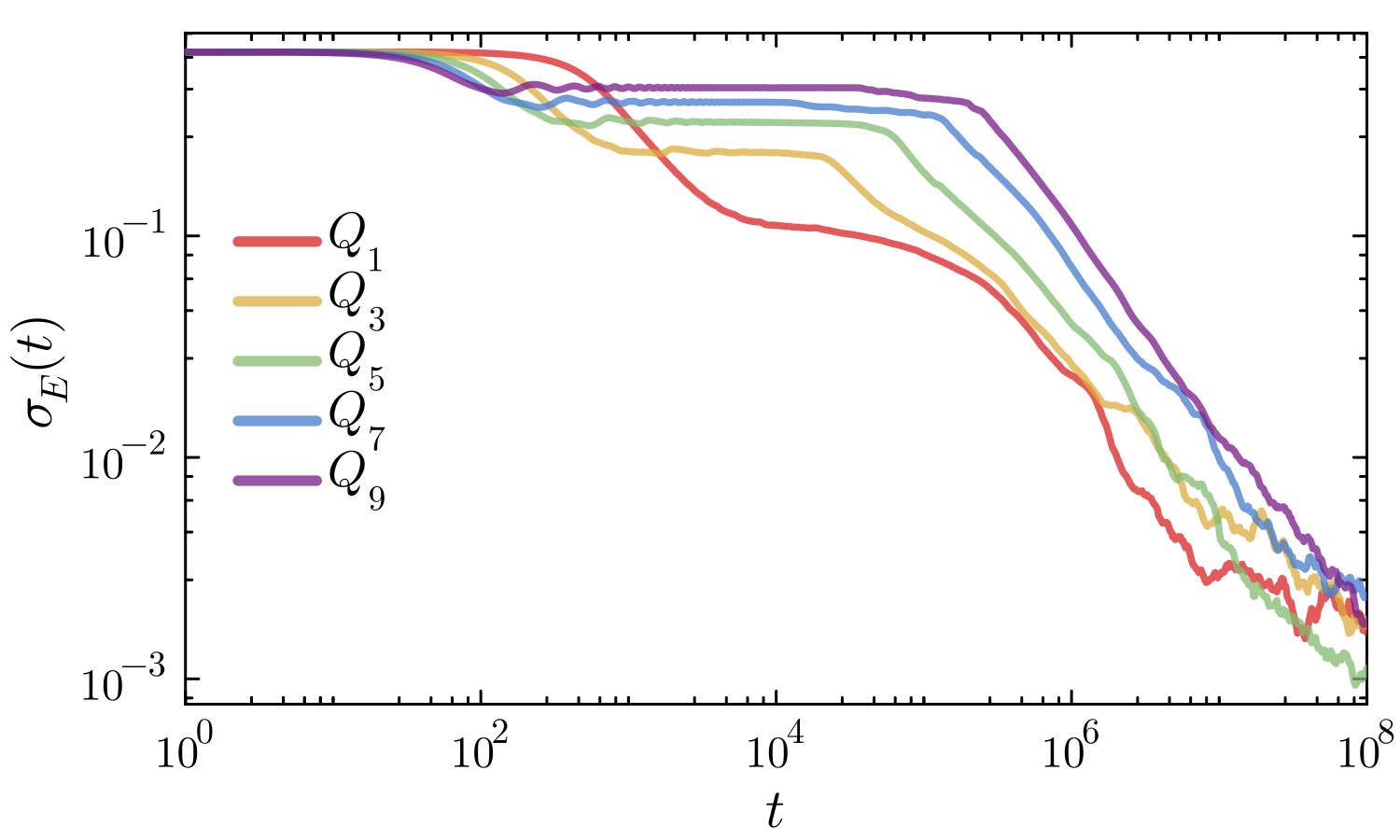}
    \caption{Standard deviation of mode energies in Fig.~\ref{fig:mode_dynamics} as a function of time for $h{\approx}0.085$. Despite keeping the energy density fixed, the prethermal regime becomes longer as a function of root number.}
    \label{fig:std_dynamics}
\end{figure}

Evidently, the longer it takes for a subset of modes to reach equipartition, the longer the prethermal regime. This is clearly seen in Fig.~\ref{fig:mode_dynamics}: The relatively obfuscated metastability associated to the $Q_1$-root seen in panel (a) is already maximized in (b), where $Q_3$ is evolved, and becomes more and more blatant as higher-excited modes are used as roots. Evidently, the random initial condition in Fig.~\ref{fig:mode_dynamics}(f) does not display metastability, since here all modes are excited at once and the slow, selection-rule mediated transferring of energy from excited modes to previously unexcited ones does not take place. Since the mean energy is a constant of motion, the prethermal regime's duration can also be visualized by plotting the time-dependent standard deviation of mode energies in \eqref{fig:mode_dynamics},  
\begin{equation}
    \sigma_E(t) = \sqrt{ \frac{1}{N-1} \sum_{j=1}^N \left( E_j(t) - h \right)^2} \, ,
\end{equation}
which will be approximately constant while the energy is still strongly localized in a subset of normal modes. This is indeed seen to be true in Fig.~\ref{fig:std_dynamics}, where it is also clear that the metastability's lifetime increases monotonically as a function of root number for this energy density. 

\subsection{Equipartition and Lyapunov times}

\begin{figure}
    \centering
    \includegraphics[width=\linewidth]{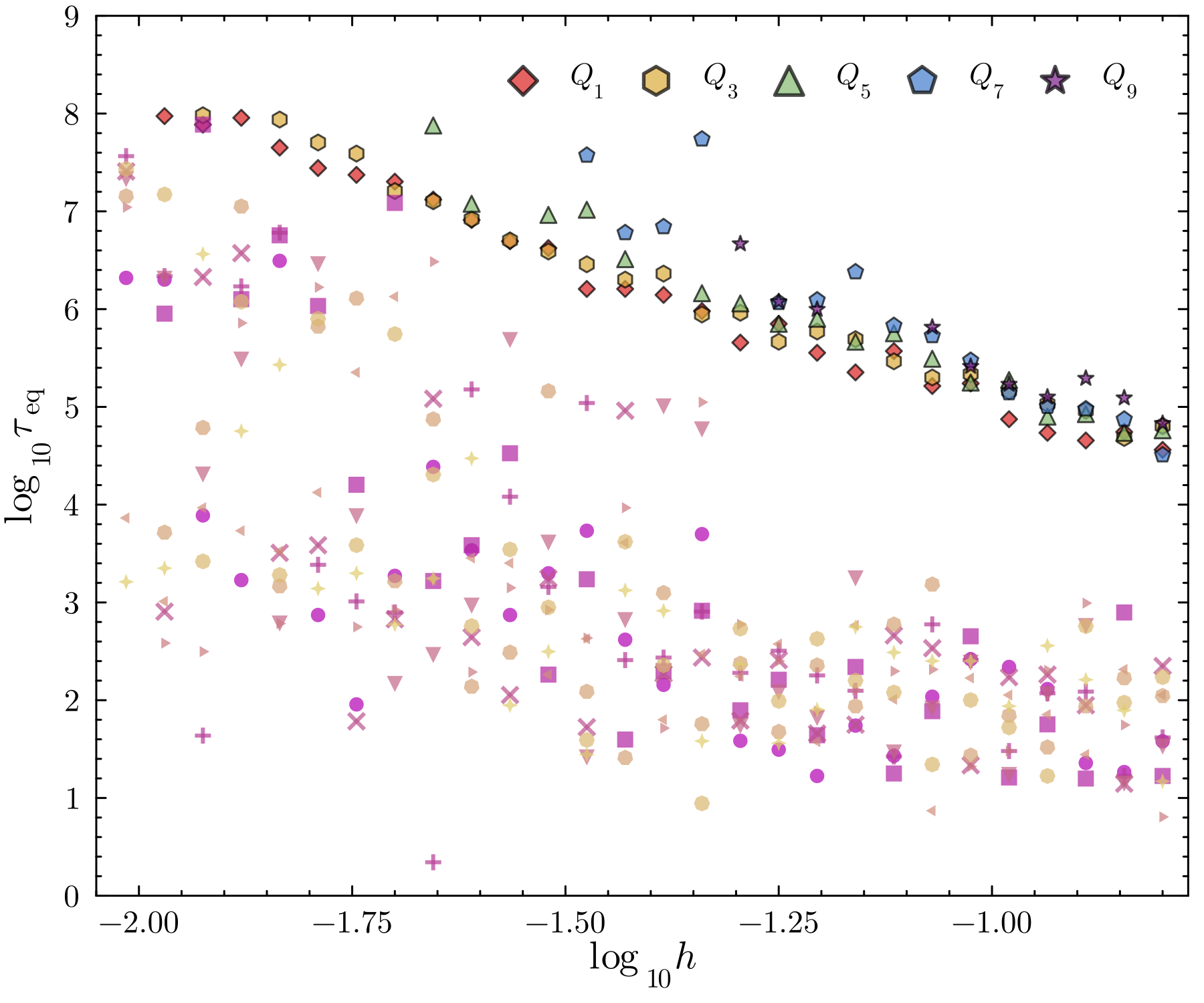}
    \caption{Equipartition times computed by evolving the first five odd roots (rainbow colors) and 280 random initial states (tones of magenta and orange) at several different energy densities in a chain with $N{=}63$ and $\alpha{=}1/4$. These data were obtained from spectral entropies computed up to $t{=}10^8$, such that the top left equipartition times, which are of $\mathcal{O}(10^8)$, are likely slightly overestimated by being of the order of the computation time. The equipartition time for $Q_1$ (red diamonds) clearly forms an upper bound that the ones computed from random initial states essentially never cross. The equipartition times computed from excited modes consistently do so, albeit slowly as a function of increasing root number.}
    \label{fig:equipartition_times}
\end{figure}

Spectral entropy \eqref{eq:spectral_entropy} is a function of the initial state, so it is no surprise that the equipartition times computed from it in Fig.~\ref{fig:mode_dynamics} depend on it. These times have no direct relation to the duration of prethermalization, except for the fact that they are necessarily longer than the length of metastable plateaus. Since random initial conditions do not, at least generally, undergo prethermal regimes, it is expected that the equipartition times of such typical initial conditions will be shorter than the ones of roots for any energy density chosen. This expectation is confirmed in Fig.~\ref{fig:equipartition_times}, where we compare equipartition times for 280 typical conditions with the ones obtained from odd roots as a function of the energy density. Clearly, the former are bounded from above by the latter, which slowly increase as a function of excitation number. 

\begin{figure}
    \centering
    \includegraphics[width=\linewidth]{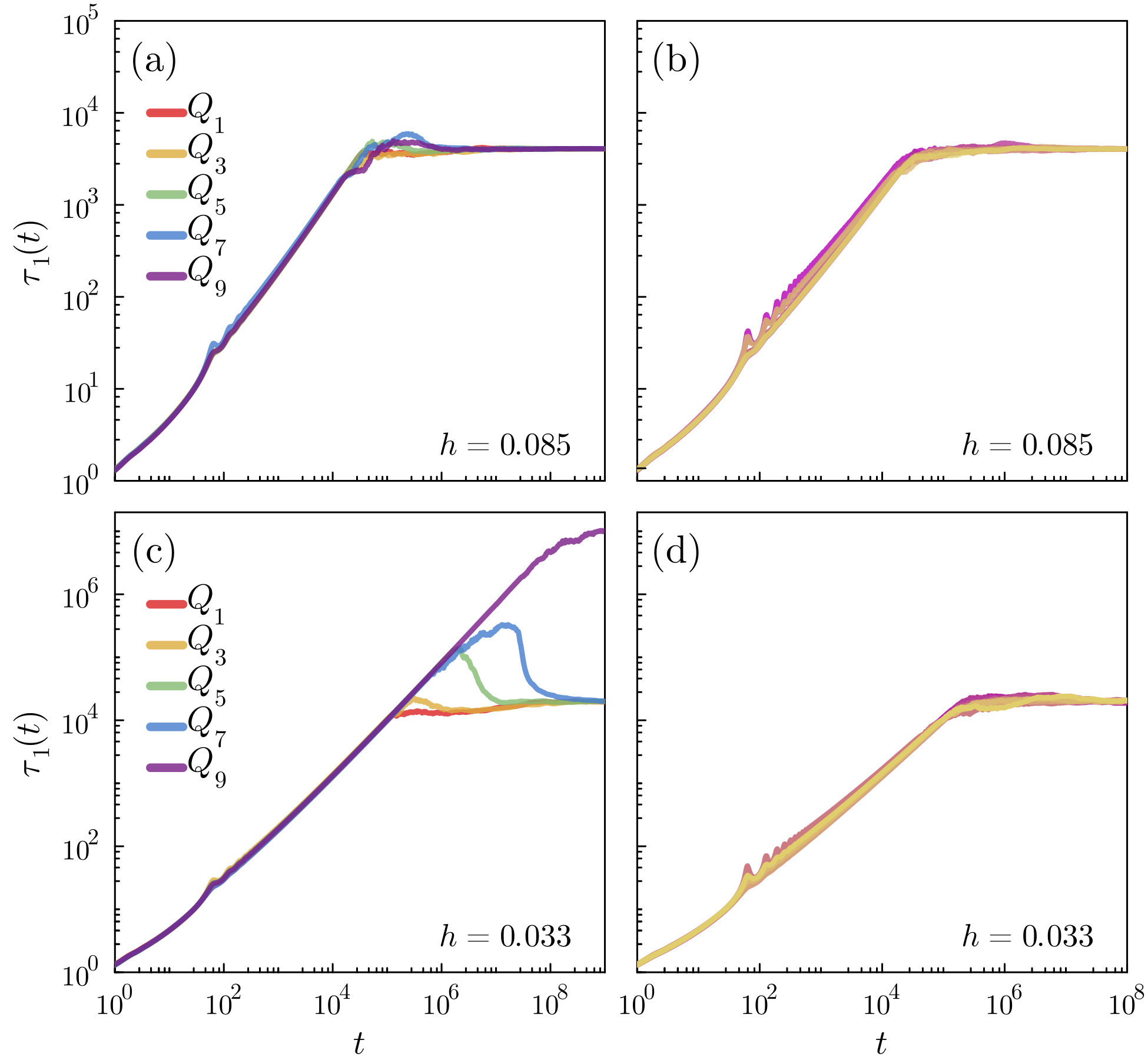}
    \caption{(a) Time-dependent Lyapunov times computed by evolving the first five odd modes of a chain with $N{=}63$ and $\alpha{=}1/4$ at a fixed energy density $h{\approx}0.085$ up to $t_f{=}10^9$. Clearly, at this energy density all modes behave similarly and indistinguishably from the random initial states shown in panel (b), for which the propagation time was set to $t_f{=}10^8$. (c) By lowering the energy density down to $h{\approx}0.033$, roots start to behave differently as a function of time when compared to the random states used in (d). The differences appear to grow with excitation number, resulting in high-order modes such $Q_9$ not converging at all within this propagation interval (which is more than enough for all other initial states).}
    \label{fig:tts}
\end{figure}

If equipartition times reach the order of the propagation time used to compute them, namely $\mathcal{O}(10^8)$, convergence can no longer be achieved for these parameters, which is clear in Fig.~\ref{fig:equipartition_times} by the presence of missing data for some values of energy density. However, the ``disappearance'' of points does not increase fully monotonically with mode excitation number, with some data displaying what could be described as ``erratic behavior''. Since dependence on initial conditions is tantamount to ergodicity breaking, investigating quantities that are proved to be invariant in any ergodic system will shed light into the source of the instabilities seen in Fig.~\ref{fig:equipartition_times}. The most important of such quantities is the MLE, $\lambda_1$, which indicates the presence of chaotic behavior and is not only invariant with respect to diffeomorphisms but also constant for ergodic dynamical systems \cite{eichhorn2001transformation}. From this exponent one can extract the Lyapunov time, $\tau_1=1/\lambda_1$, which measures the time a vector tangent to a trajectory takes to align with the maximally chaotic, one-dimensional submanifold in the tangent bundle \cite{oseledec1968multiplicative, eckmann1985ergodic, ginelli2007characterizing}. Since the Lyapunov time is computed from the divergence of two initially close trajectories, it takes a finite time to converge and it is often useful to observe its evolution as a function of time, as done in Fig.~\ref{fig:tts}. 

Panels (a) and (b) of Fig.~\ref{fig:tts} display the evolution of $\tau_1(t)$ extracted from propagating odd roots and random initial conditions for an energy density of $h{\approx}0.085$. Not only is convergence achieved for all initial states employed, but the time-dependent portrait of both typical and atypical initial conditions is identical. Upon decreasing the energy density, however, Fig.~\ref{fig:tts}(c) shows that atypical initial conditions start behaving very differently not only with respect to typical ones, but also among themselves. While the Lyapunov times obtained from $Q_1$ and $Q_3$ are similar to the ones of the random initial conditions in Fig.~\ref{fig:tts}(d), the data obtained from evolving $Q_5$ is seen to start forming a ``belly'' before convergence. Thus, it shows that $Q_5$ lies in a region of phase space that takes longer to align with the maximally chaotic direction, similarly to the trapping times described in \cite{casetti1997fermi}. The difference here is that we observe longer trapping times at fixed energy density as the root number increases. 

\begin{figure}
    \centering
    \includegraphics[width=\linewidth]{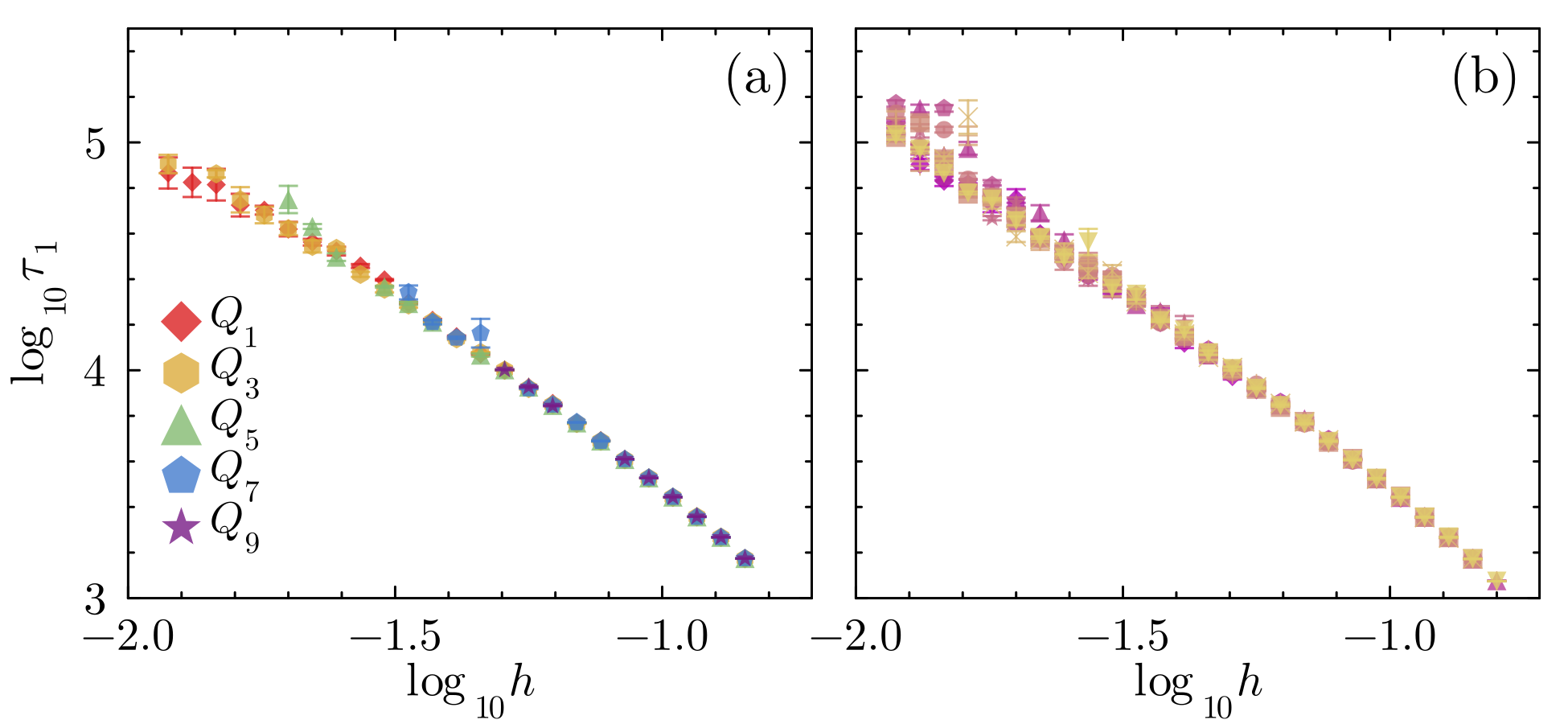}
    \caption{(a) Lyapunov times computed by evolving the first five odd roots (rainbow colors) of a  chain with $N{=}63$ and $\alpha{=}1/4$ at several different energy densities. Propagation time used was $t_f{=}10^9$, for which $\tau_1$'s computed by evolving $Q_1$ converge well for all energy densities used. Other modes, however, did not converge for the whole range of energy densities employed here, requiring longer and unpractical propagation times (see text for explanation). (b) Lyapunov times computed by evolving 10 random initial states at the same energy densities as panel (a). Propagation times in this case were only $t_f{=}10^7$, which is enough to obtain converged results starting from typical initial conditions. Error bars in all plots are the standard deviation of the corresponding $\tau_1(t)$ between $t_f$ and $t_f/10$.}
    \label{fig:t1s}
\end{figure}

\begin{figure*}[t!]
    \includegraphics[width=\linewidth]{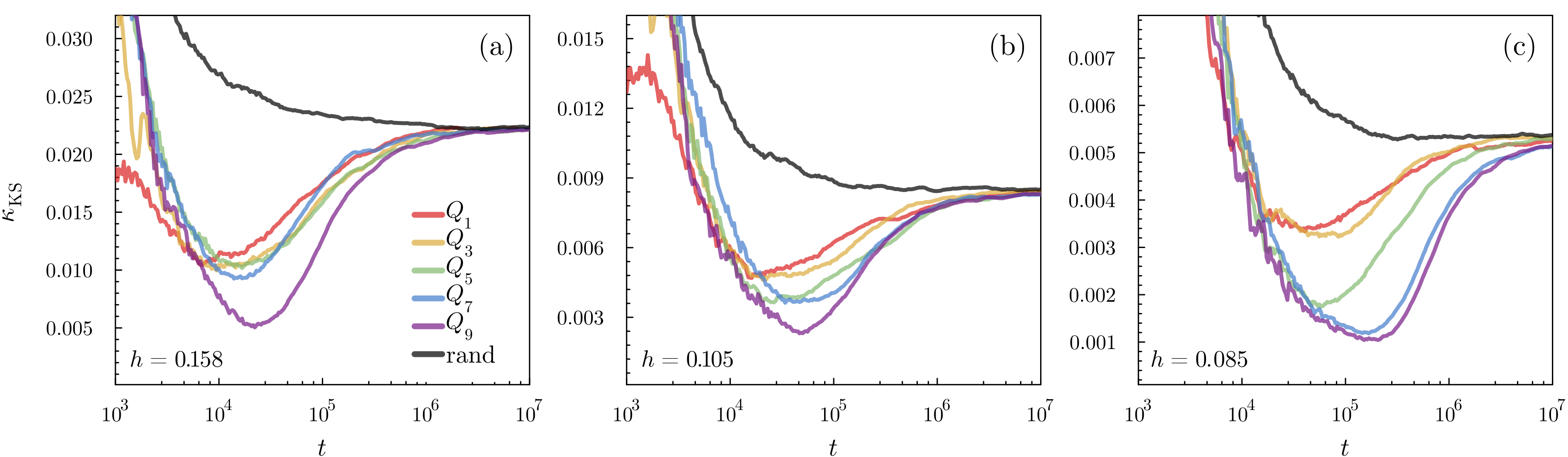}
    \caption{Time-dependent Kolmogorov-Sinai entropy in an $\alpha$-FPUT chain with $N{=}63$ and $\alpha{=}1/4$ computed from roots (rainbow colors) and from a random state (black). Each panel is computed for initial states with decreasing energy densities, $h$, shown in the left corner.}
    \label{fig:kappas}
\end{figure*}

By evolving even higher excited roots such as $Q_7$, Fig.~\ref{fig:tts}(c) shows that the ``belly'' increases when compared to the one of $Q_5$, such that $Q_7$ is trapped for a time one order of magnitude longer. Extrapolating to $Q_9$ this transient trapping regime, which is a manifestation of prethermalization in an invariant quantity and therefore \emph{must} disappear for sufficiently long times, ends up being longer than the propagation time and does not converge in our computations. The reason for the longer transient times as we go up from $Q_1$ to $Q_9$ is that the higher the excitation number, the smaller the subspace of mode space to which dynamics is approximately restricted during the prethermal regime and the higher the mode energies, as discussed in Sec.~\ref{subsec:graphs}. Indeed, Fig.~\ref{fig:mode_energies_per_site} shows that a large fraction of the energy is still concentrated in multiples of the initial excitation number at equipartition time, which as seen by comparing Figs.~\ref{fig:tts} and \ref{fig:t1s} is always longer than Lyapunov time when dealing with atypical initial conditions. Thus, the trapping time will increase with the excitation number, which is also a consequence of higher-excited modes lying closer to q-breathers.

Fig.~\ref{fig:t1s}, which shows the final (converged) Lyapunov times as a function of energy density for typical and atypical initial conditions, is clear evidence that the system we are dealing with is numerically ergodic: The times are the same, no matter what initial conditions are chosen (although the larger error bars and spread for $h{\approx}10^{-2}$ show that $t_f{=}10^{9}$ is barely long enough to reach numerical ergodicity for the smallest values of $h$ used here). This does not mean, as discussed before, that all data for roots converges, with the main reason for failure being the presence of extremely long prethermal regimes as seen in Fig.~\ref{fig:tts}. 

\subsection{Kolmogorov-Sinai entropies}

Given the imprint prethermalization has in a dynamical invariant such as the Lyapunov time, in Fig.~\ref{fig:kappas} we display the time evolution of one of the most meaningful, yet computationally expensive, invariant quantities in dynamical systems, namely the Kolmogorov-Sinai entropy, $\kappa_\mathrm{KS}$. This quantity, just like the Lyapunov time and the associated maximal exponent, displays a clear dip when computed by propagating roots, while for a typical initial condition it approaches its converged value from above. Since $\kappa_\mathrm{KS}$ is the sum of all positive Lyapunov exponents, what Fig.~\ref{fig:kappas} shows is that the prethermal regime is present in all phase-space directions, not only the maximal one associated to the MLE. Upon decreasing the energy density we also see a tendency of a larger dip for higher-excited modes, in accordance with our expectation that the prethermal regime is more pronounced for higher excitation numbers. It is worth noting that the convergence of Kolmogorov–Sinai entropies to the same asymptotic value likely constitutes the most compelling numerical evidence of ergodicity attainable in simulations.

\section{Discussion}\label{sec:discussion}

The root chosen in the original FPUT experiment and most subsequent investigations of the FPUT paradox was the fundamental mode. In essence, and as previously shown by comparisons with the nearby and integrable Toda chain \cite{casetti1997fermi}, the proximity of the fundamental mode to a stationary state of the FPUT chain ends up trapping it in a near-regular portion of phase space for a finite time. We have shown here that higher-excited roots are trapped for even longer than the fundamental mode in such near-regular regions, even though all other parameters in the system are held constant.

\begin{figure}[t!]
    \centering
    \includegraphics[width=\linewidth]{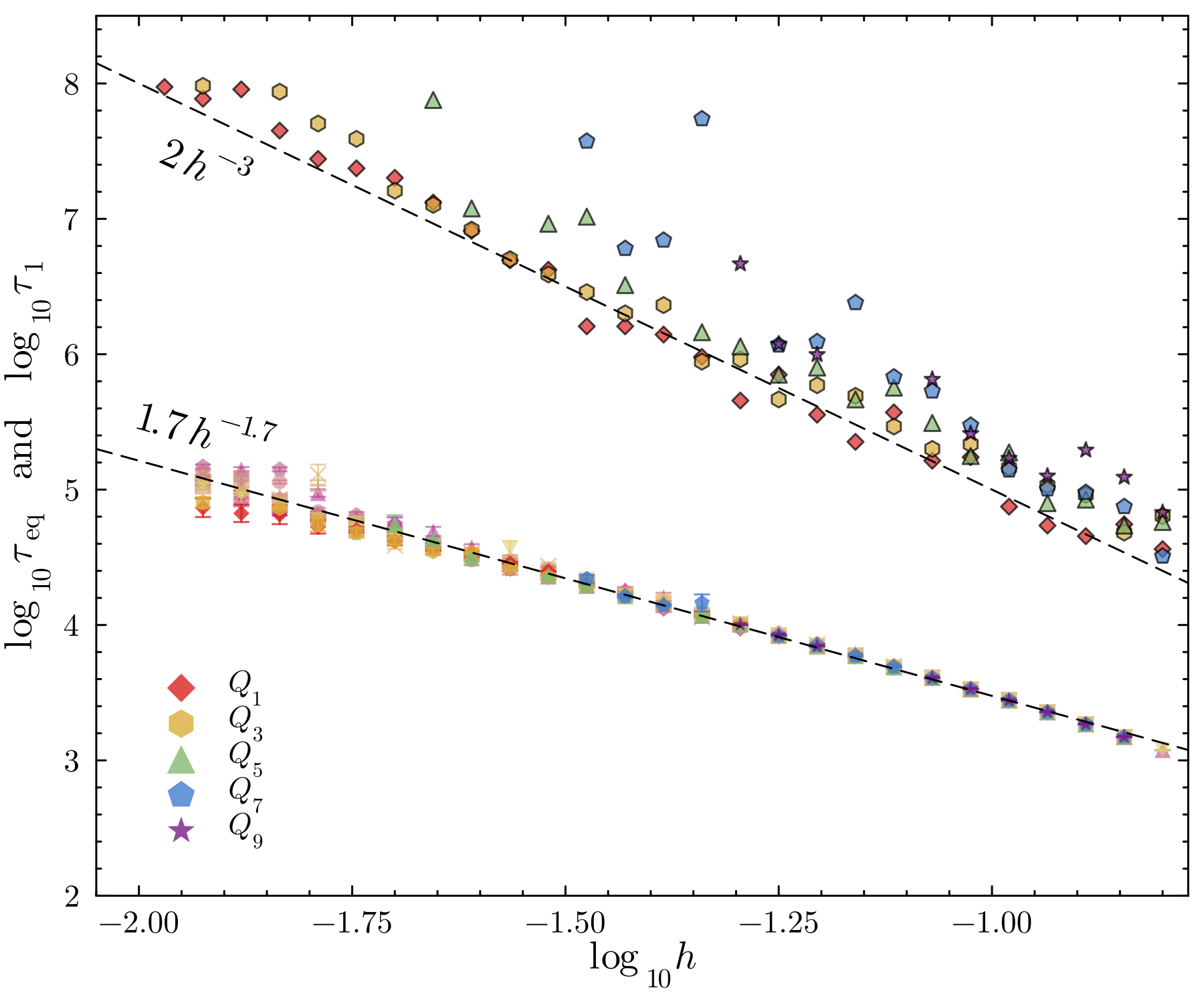}
    \caption{Comparison of Lyapunov times in Fig.~\ref{fig:t1s}, where here both panels are superposed, and its corresponding linear fit. The equipartition times for root modes in Fig.~\ref{fig:equipartition_times} are also displayed, together with a linear fit for those of $Q_1$, showing that they are orders of magnitude longer than the Lyapunov times for all energy densities.} 
    \label{fig:time_comparisons}
\end{figure}

Traditionally, the Lyapunov time is considered to characterize alignment along the maximally chaotic direction in a decomposition of the tangent space known as Oseledec splitting, which is covariant with respect to the Hamiltonian flow \cite{oseledec1968multiplicative, ginelli2007characterizing}. Evidently, such decomposition is meaningless in the case of a near-regular trajectory, and it is expected that the transient regime of $\lambda_1(t)$ will be different depending on how close the initial state is to a near-integrable region. However, the Lyapunov time itself is an invariant quantity in ergodic systems, such that at the end all initial states must provide the same value for the converged MLE (that is, the asymptotic limit of $\lambda_1(t))$. This was verified in Figs.~\ref{fig:tts} and \ref{fig:t1s}. Thus, the time needed to escape a region of near-integrability is more related to the equipartition time than to Lyapunov time, since we expect mode space to be fully covered only once a trajectory starting in a root becomes sufficiently chaotic. The Lyapunov time itself then carries absolutely no information on prethermalization, although its time-evolution certainly does. This leads to a reinterpretation of $\tau_1$ as the time a \emph{typical} initial condition takes to align with the maximally chaotic covariant direction in tangent space, since atypical ones might not align with it at all. This results in the Lyapunov time being always shorter than the equipartition time of a root, as seen by plotting them together in Fig.~\ref{fig:time_comparisons}.

The only time scale longer than the equipartition time of roots appears to be the convergence time of their time-dependent Kolmogorov-Sinai entropies, $\kappa_\mathrm{KS}(t)$. Like $\lambda_1(t)$, this quantity will only converge once the trajectory leaves a near-regular region, but now alignment with respect to a large fraction of covariant Lyapunov directions is required instead of only the maximal one. Since several directions are associated to small Lyapunov exponents it is not necessary to align with \emph{all} of them to have a well-converged result, but Fig.~\ref{fig:kappas}(c) shows that the convergence times are, nevertheless, longer than the equipartition times shown in Fig.~\ref{fig:mode_dynamics}. Thus, the inescapable conclusion is that equipartition is reached before the trajectory has fully explored the chaotic sea.

It would be interesting to attempt a computation of the Lyapunov time of a submanifold, i.e., calculate $\tau_1$ starting from a nonthermal mode, e.g., $Q_2$, and see how this exponent relates to the true $\tau_1$ obtained from a thermal initial condition. Unfortunately this type of numerical investigation is hard, if not impossible, to carry on. This is due to the fact that the bushes associated to nonthermal modes in $\alpha$-FPUT chains are unstable and will not be preserved in long-time evolution no matter which numerical integrator is chosen \cite{chechin2002bushes}. 
Indeed, if the chosen root is nonthermal, the time evolution of the mode energies reveals that the initial excitation eventually escapes the bush and spreads throughout mode space, as numerical integration does not preserve the discrete symmetries of bushes.
This might be behind the observation that the evolution of nonthermal modes results in approximately the same Lyapunov times as thermal ones \cite{casetti1997fermi}, which is surprising given that bush dynamics takes place only in a submanifold of mode space. 
Nevertheless, it is unclear how ergodicity in mode space relates to that of phase space, such that it is possible that a partial covering of mode space by a large enough bush is enough to resolve the ``true'' MLE. Besides, it might even happen that dynamics in phase space is ergodic while that in mode space is not, since ergodicity always depends on the observables chosen \cite{baldovin2021statistical} unless it is tracked by means of invariant quantities \cite{lando2023thermalization}. 
At present, the authors are unaware of a special type of integrator that is capable of preserving bushes of excitations, such that studies regarding dynamical properties of chaotic subspaces cannot be performed in a numerically meaningful fashion. A consequence of this is that the energy in Fig.~\ref{fig:sequences}(a), which is correctly localized in the bush associated to $Q_3$, will eventually spread to all modes for long enough times. Nevertheless, the way ``equipartition'' takes place in this case is pathological and clearly traceable to numerical errors, as can be seen in Appendix \ref{app:num_errors}.

\section{Conclusion}\label{sec:conclusion}

We have extended previous studies on the cubic Fermi-Pasta-Ulam-Tsingou ($\alpha$-FPUT) model by investigating the consequences of evolving higher-excited normal modes instead of the fundamental. The evolution of such atypical initial conditions was also compared to that obtained from typical (random) ones. We provided an explicit and simple way of ordering excited modes perturbatively in $\alpha$ when starting from a single normal mode (the root), and showed that higher-excited roots take longer to cover mode space. Thorough numerical investigations based on computing Lyapunov exponents and Kolmogorov-Sinai entropies showed that even invariant quantities carry imprints of the atypical initial states used to compute them in their transient regime, despite converging to the correct invariant value for long times. The prethermal dynamics taking place before convergence is confirmed to be a consequence of some select roots lying close to stationary states also in the $\alpha$-FPUT system, and therefore being trapped in near-regular regions. The escape from such regions is what puts an end to the paradigmatic metastable plateaus in mode energies, whose end marks the moment invariant quantities computed from roots start to converge. 

\acknowledgements

This research was supported by the Institute for Basic Science
through project IBS-R024-D1.

\appendix

\section{Numerical instability of bushes}\label{app:num_errors}

\begin{figure}
    \centering
    \includegraphics[width=\linewidth]{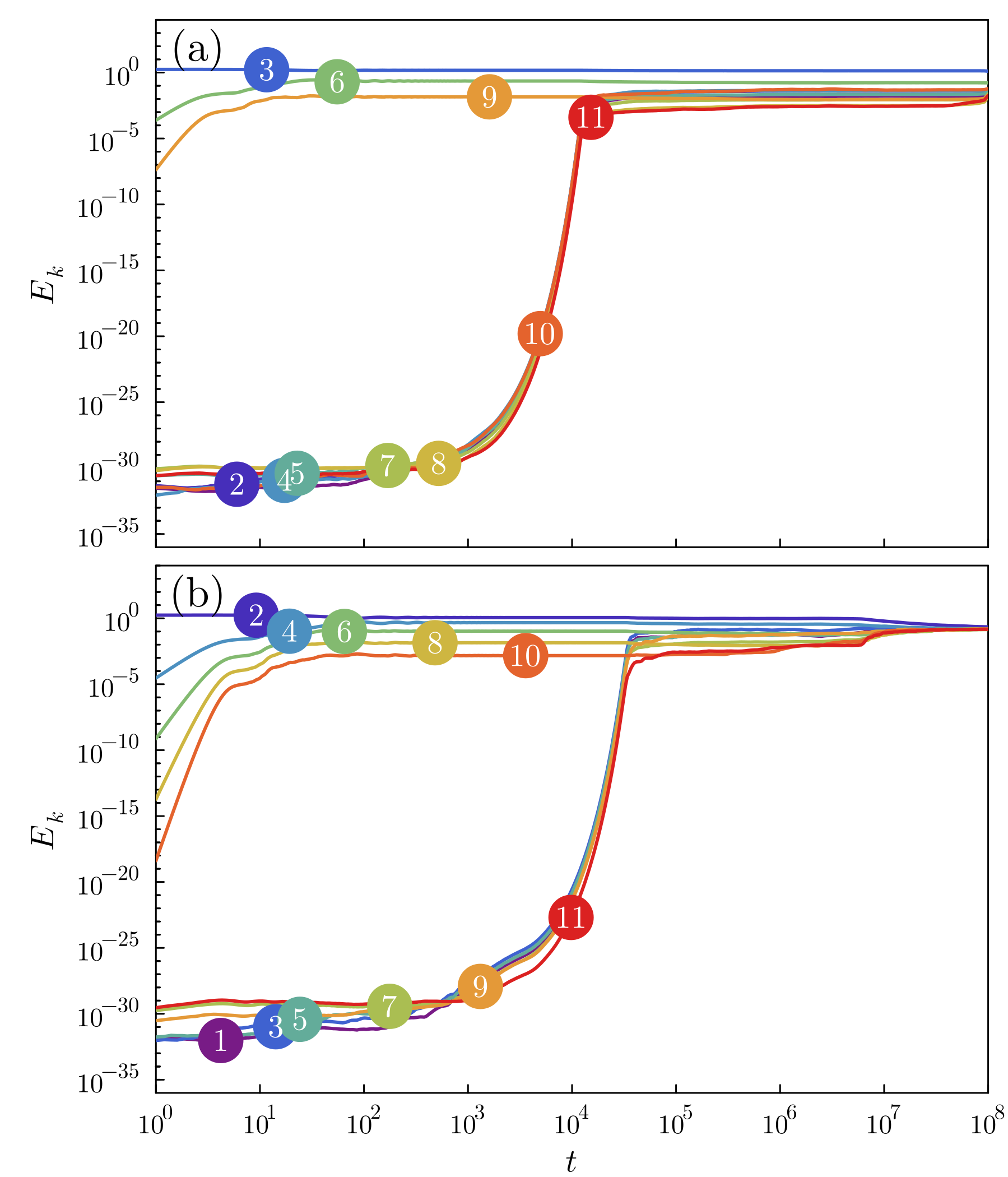}
    \caption{Examples of error-induced equipartition due to numerical instability of bushes in a chain with $N{=}11$, $h{\approx}0.16$ and $\alpha{=}1/4$. (a) Same as Fig.~\ref{fig:sequences}(a) but for longer times. (b) The evolution of $Q_2$, which as all even roots is nonthermal. In both panels, states that are proven to be nonthermal end up reaching equipartition due to numerical errors. Panels (a) and (b) use different symplectic integrators (second and fourth orders), showing that the phenomenon happens independently of the solver, although it happens at different times depending on the time step (attesting for its numerical source).}
    \label{fig:numerical_errors}
\end{figure}

The bushes associated with nonthermal roots in $\alpha$-FPUT chains obey discrete symmetries that are not respected by the integrators, being therefore unstable in numerical simulations. The main consequence of this instability is that energy, which should only be shared among modes in the bush and never leave for any finite time, ends up exciting other modes due to numerical errors. In Fig.~\ref{fig:numerical_errors} we provide two examples of this behavior when evolving provenly nonthermal modes: the roots $Q_3$ and $Q_2$ in a $N=11$ chain. The first case is essentially a continuation of Fig.~\ref{fig:sequences} for much longer times, until one can see an anomalous and simultaneous jump in the energy of modes that should not be excited starting from $Q_3$; the root $Q_2$, on the other hand, is even and therefore always nonthermal, undergoing a very similar pathological jump and reaching equipartition in Fig.~\ref{fig:numerical_errors} exclusively due to numerical errors.

\bibliography{bib, sergejflach}

\begin{thebibliography}{47}%
\makeatletter
\providecommand \@ifxundefined [1]{%
 \@ifx{#1\undefined}
}%
\providecommand \@ifnum [1]{%
 \ifnum #1\expandafter \@firstoftwo
 \else \expandafter \@secondoftwo
 \fi
}%
\providecommand \@ifx [1]{%
 \ifx #1\expandafter \@firstoftwo
 \else \expandafter \@secondoftwo
 \fi
}%
\providecommand \natexlab [1]{#1}%
\providecommand \enquote  [1]{``#1''}%
\providecommand \bibnamefont  [1]{#1}%
\providecommand \bibfnamefont [1]{#1}%
\providecommand \citenamefont [1]{#1}%
\providecommand \href@noop [0]{\@secondoftwo}%
\providecommand \href [0]{\begingroup \@sanitize@url \@href}%
\providecommand \@href[1]{\@@startlink{#1}\@@href}%
\providecommand \@@href[1]{\endgroup#1\@@endlink}%
\providecommand \@sanitize@url [0]{\catcode `\\12\catcode `\$12\catcode `\&12\catcode `\#12\catcode `\^12\catcode `\_12\catcode `\%12\relax}%
\providecommand \@@startlink[1]{}%
\providecommand \@@endlink[0]{}%
\providecommand \url  [0]{\begingroup\@sanitize@url \@url }%
\providecommand \@url [1]{\endgroup\@href {#1}{\urlprefix }}%
\providecommand \urlprefix  [0]{URL }%
\providecommand \Eprint [0]{\href }%
\providecommand \doibase [0]{https://doi.org/}%
\providecommand \selectlanguage [0]{\@gobble}%
\providecommand \bibinfo  [0]{\@secondoftwo}%
\providecommand \bibfield  [0]{\@secondoftwo}%
\providecommand \translation [1]{[#1]}%
\providecommand \BibitemOpen [0]{}%
\providecommand \bibitemStop [0]{}%
\providecommand \bibitemNoStop [0]{.\EOS\space}%
\providecommand \EOS [0]{\spacefactor3000\relax}%
\providecommand \BibitemShut  [1]{\csname bibitem#1\endcsname}%
\let\auto@bib@innerbib\@empty
\bibitem [{\citenamefont {Fermi}\ \emph {et~al.}(1955)\citenamefont {Fermi}, \citenamefont {Pasta}, \citenamefont {Ulam},\ and\ \citenamefont {Tsingou}}]{fermi1955studies}%
  \BibitemOpen
  \bibfield  {author} {\bibinfo {author} {\bibfnamefont {E.}~\bibnamefont {Fermi}}, \bibinfo {author} {\bibfnamefont {J.}~\bibnamefont {Pasta}}, \bibinfo {author} {\bibfnamefont {S.}~\bibnamefont {Ulam}},\ and\ \bibinfo {author} {\bibfnamefont {M.}~\bibnamefont {Tsingou}},\ }\href@noop {} {\emph {\bibinfo {title} {Studies of nonlinear problems. I}}},\ \bibinfo {type} {Tech. Rep.}\ \bibinfo {number} {LA-1940}\ (\bibinfo  {institution} {Los Alamos Scientific Lab., N. Mex.},\ \bibinfo {year} {1955})\BibitemShut {NoStop}%
\bibitem [{\citenamefont {Ford}(1992)}]{ford1992fermi}%
  \BibitemOpen
  \bibfield  {author} {\bibinfo {author} {\bibfnamefont {J.}~\bibnamefont {Ford}},\ }\bibfield  {title} {\bibinfo {title} {The {F}ermi-{P}asta-{U}lam problem: paradox turns discovery},\ }\href@noop {} {\bibfield  {journal} {\bibinfo  {journal} {Physics Reports}\ }\textbf {\bibinfo {volume} {213}},\ \bibinfo {pages} {271} (\bibinfo {year} {1992})}\BibitemShut {NoStop}%
\bibitem [{\citenamefont {Weissert}(1997)}]{fpugen}%
  \BibitemOpen
  \bibfield  {author} {\bibinfo {author} {\bibfnamefont {T.~P.}\ \bibnamefont {Weissert}},\ }\href@noop {} {\emph {\bibinfo {title} {{The Genesis of Simulation in Dynamics: Pursuing the Fermi-Pasta-Ulam Problem}}}}\ (\bibinfo  {publisher} {Springer-Verlag},\ \bibinfo {address} {New York, NY},\ \bibinfo {year} {1997})\BibitemShut {NoStop}%
\bibitem [{\citenamefont {Gallavotti}(2007)}]{gallavotti2007fermi}%
  \BibitemOpen
  \bibfield  {author} {\bibinfo {author} {\bibfnamefont {G.}~\bibnamefont {Gallavotti}},\ }\href@noop {} {\emph {\bibinfo {title} {The {F}ermi-{P}asta-{U}lam problem: a status report}}},\ Vol.\ \bibinfo {volume} {728}\ (\bibinfo  {publisher} {Springer Science \& Business Media},\ \bibinfo {year} {2007})\BibitemShut {NoStop}%
\bibitem [{\citenamefont {Kolmogorov}(1954)}]{kolmogorov1954conservation}%
  \BibitemOpen
  \bibfield  {author} {\bibinfo {author} {\bibfnamefont {A.~N.}\ \bibnamefont {Kolmogorov}},\ }\bibfield  {title} {\bibinfo {title} {On conservation of conditionally periodic motions for a small change in {H}amilton's function},\ }\href@noop {} {\bibfield  {journal} {\bibinfo  {journal} {Doklady Akademii Nauk SSSR}\ }\textbf {\bibinfo {volume} {98}},\ \bibinfo {pages} {527} (\bibinfo {year} {1954})},\ \bibinfo {note} {translated into English in Lecture Notes in Physics, vol. 93, 1979, Springer}\BibitemShut {NoStop}%
\bibitem [{\citenamefont {Arnold}(1963)}]{arnold1963small}%
  \BibitemOpen
  \bibfield  {author} {\bibinfo {author} {\bibfnamefont {V.~I.}\ \bibnamefont {Arnold}},\ }\bibfield  {title} {\bibinfo {title} {Small denominators. {I}. mappings of the circle onto itself},\ }\href@noop {} {\bibfield  {journal} {\bibinfo  {journal} {Izvestiya Akademii Nauk SSSR. Seriya Matematicheskaya}\ }\textbf {\bibinfo {volume} {25}},\ \bibinfo {pages} {21} (\bibinfo {year} {1963})},\ \bibinfo {note} {translated into English in American Mathematical Society Translations, Series 2, vol. 46, 1965, pp. 213-284}\BibitemShut {NoStop}%
\bibitem [{\citenamefont {Moser}(1962)}]{moser1962invariant}%
  \BibitemOpen
  \bibfield  {author} {\bibinfo {author} {\bibfnamefont {J.}~\bibnamefont {Moser}},\ }\bibfield  {title} {\bibinfo {title} {On invariant curves of area-preserving mappings of an annulus},\ }\href@noop {} {\bibfield  {journal} {\bibinfo  {journal} {Nachrichten der Akademie der Wissenschaften in G{\"o}ttingen. II. Mathematisch-Physikalische Klasse}\ }\textbf {\bibinfo {volume} {1962}},\ \bibinfo {pages} {1} (\bibinfo {year} {1962})},\ \bibinfo {note} {reprinted in J. Moser, Stable and Random Motions in Dynamical Systems, Princeton University Press, 1973, pp. 89-112}\BibitemShut {NoStop}%
\bibitem [{\citenamefont {Chirikov}(1979)}]{chirikov1979universal}%
  \BibitemOpen
  \bibfield  {author} {\bibinfo {author} {\bibfnamefont {B.~V.}\ \bibnamefont {Chirikov}},\ }\bibfield  {title} {\bibinfo {title} {A universal instability of many-dimensional oscillator systems},\ }\href@noop {} {\bibfield  {journal} {\bibinfo  {journal} {Physics reports}\ }\textbf {\bibinfo {volume} {52}},\ \bibinfo {pages} {263} (\bibinfo {year} {1979})}\BibitemShut {NoStop}%
\bibitem [{\citenamefont {Birkhoff}(1927)}]{birkhoff1927periodic}%
  \BibitemOpen
  \bibfield  {author} {\bibinfo {author} {\bibfnamefont {G.~D.}\ \bibnamefont {Birkhoff}},\ }\bibfield  {title} {\bibinfo {title} {On the periodic motions of dynamical systems},\ }\href {https://doi.org/10.1007/BF02559588} {\bibfield  {journal} {\bibinfo  {journal} {Acta Mathematica}\ }\textbf {\bibinfo {volume} {50}},\ \bibinfo {pages} {359} (\bibinfo {year} {1927})}\BibitemShut {NoStop}%
\bibitem [{\citenamefont {Gustavson}(1966)}]{gustavson1966formal}%
  \BibitemOpen
  \bibfield  {author} {\bibinfo {author} {\bibfnamefont {F.~G.}\ \bibnamefont {Gustavson}},\ }\bibfield  {title} {\bibinfo {title} {On constructing formal integrals of a {H}amiltonian system near an equilibrium point},\ }\href {https://doi.org/10.1086/109855} {\bibfield  {journal} {\bibinfo  {journal} {Astronomical Journal}\ }\textbf {\bibinfo {volume} {71}},\ \bibinfo {pages} {670} (\bibinfo {year} {1966})}\BibitemShut {NoStop}%
\bibitem [{\citenamefont {Ozorio~de Almeida}(1990)}]{ozorio1990hamiltonian}%
  \BibitemOpen
  \bibfield  {author} {\bibinfo {author} {\bibfnamefont {A.~M.}\ \bibnamefont {Ozorio~de Almeida}},\ }\href {https://doi.org/10.1017/CBO9780511569586} {\emph {\bibinfo {title} {Hamiltonian Systems: Chaos and Quantization}}},\ Cambridge Monographs on Mathematical Physics\ (\bibinfo  {publisher} {Cambridge University Press},\ \bibinfo {address} {Cambridge},\ \bibinfo {year} {1990})\BibitemShut {NoStop}%
\bibitem [{\citenamefont {Murdock}(2003)}]{murdock2003normal}%
  \BibitemOpen
  \bibfield  {author} {\bibinfo {author} {\bibfnamefont {J.~A.}\ \bibnamefont {Murdock}},\ }\href {https://doi.org/10.1007/b97424} {\emph {\bibinfo {title} {Normal Forms and Unfoldings for Local Dynamical Systems}}},\ \bibinfo {series} {Applied Mathematical Sciences}, Vol.\ \bibinfo {volume} {182}\ (\bibinfo  {publisher} {Springer},\ \bibinfo {address} {New York},\ \bibinfo {year} {2003})\BibitemShut {NoStop}%
\bibitem [{\citenamefont {Wiggins}(2003)}]{wiggins2003introduction}%
  \BibitemOpen
  \bibfield  {author} {\bibinfo {author} {\bibfnamefont {S.}~\bibnamefont {Wiggins}},\ }\href {https://doi.org/10.1007/b97481} {\emph {\bibinfo {title} {Introduction to Applied Nonlinear Dynamical Systems and Chaos}}},\ \bibinfo {edition} {2nd}\ ed.,\ \bibinfo {series} {Texts in Applied Mathematics}, Vol.~\bibinfo {volume} {2}\ (\bibinfo  {publisher} {Springer},\ \bibinfo {address} {New York},\ \bibinfo {year} {2003})\BibitemShut {NoStop}%
\bibitem [{\citenamefont {Pace}\ and\ \citenamefont {Campbell}(2019)}]{pace2019behavior}%
  \BibitemOpen
  \bibfield  {author} {\bibinfo {author} {\bibfnamefont {S.~D.}\ \bibnamefont {Pace}}\ and\ \bibinfo {author} {\bibfnamefont {D.~K.}\ \bibnamefont {Campbell}},\ }\bibfield  {title} {\bibinfo {title} {{Behavior and breakdown of higher-order Fermi-Pasta-Ulam-Tsingou recurrences}},\ }\href@noop {} {\bibfield  {journal} {\bibinfo  {journal} {Chaos: An Interdisciplinary Journal of Nonlinear Science}\ }\textbf {\bibinfo {volume} {29}},\ \bibinfo {pages} {023132} (\bibinfo {year} {2019})}\BibitemShut {NoStop}%
\bibitem [{\citenamefont {Poggi}\ and\ \citenamefont {Ruffo}(1997)}]{poggi1997exact}%
  \BibitemOpen
  \bibfield  {author} {\bibinfo {author} {\bibfnamefont {P.}~\bibnamefont {Poggi}}\ and\ \bibinfo {author} {\bibfnamefont {S.}~\bibnamefont {Ruffo}},\ }\bibfield  {title} {\bibinfo {title} {Exact solutions in the {FPU} oscillator chain},\ }\href@noop {} {\bibfield  {journal} {\bibinfo  {journal} {Physica D: Nonlinear Phenomena}\ }\textbf {\bibinfo {volume} {103}},\ \bibinfo {pages} {251} (\bibinfo {year} {1997})}\BibitemShut {NoStop}%
\bibitem [{\citenamefont {Berman}\ and\ \citenamefont {Izrailev}(2005)}]{berman2005fermi}%
  \BibitemOpen
  \bibfield  {author} {\bibinfo {author} {\bibfnamefont {G.}~\bibnamefont {Berman}}\ and\ \bibinfo {author} {\bibfnamefont {F.}~\bibnamefont {Izrailev}},\ }\bibfield  {title} {\bibinfo {title} {The {F}ermi--{P}asta--{U}lam problem: fifty years of progress},\ }\href@noop {} {\bibfield  {journal} {\bibinfo  {journal} {Chaos: An Interdisciplinary Journal of Nonlinear Science}\ }\textbf {\bibinfo {volume} {15}} (\bibinfo {year} {2005})}\BibitemShut {NoStop}%
\bibitem [{\citenamefont {Campbell}\ \emph {et~al.}(2005)\citenamefont {Campbell}, \citenamefont {Rosenau},\ and\ \citenamefont {Zaslavsky}}]{campbell2005introduction}%
  \BibitemOpen
  \bibfield  {author} {\bibinfo {author} {\bibfnamefont {D.~K.}\ \bibnamefont {Campbell}}, \bibinfo {author} {\bibfnamefont {P.}~\bibnamefont {Rosenau}},\ and\ \bibinfo {author} {\bibfnamefont {G.~M.}\ \bibnamefont {Zaslavsky}},\ }\bibfield  {title} {\bibinfo {title} {Introduction: the {F}ermi--{P}asta--{U}lam problem—the first fifty years},\ }\href@noop {} {\bibfield  {journal} {\bibinfo  {journal} {Chaos: An Interdisciplinary Journal of Nonlinear Science}\ }\textbf {\bibinfo {volume} {15}} (\bibinfo {year} {2005})}\BibitemShut {NoStop}%
\bibitem [{\citenamefont {Benettin}\ \emph {et~al.}(2008)\citenamefont {Benettin}, \citenamefont {Carati}, \citenamefont {Galgani},\ and\ \citenamefont {Giorgilli}}]{benettin2008fermi}%
  \BibitemOpen
  \bibfield  {author} {\bibinfo {author} {\bibfnamefont {G.}~\bibnamefont {Benettin}}, \bibinfo {author} {\bibfnamefont {A.}~\bibnamefont {Carati}}, \bibinfo {author} {\bibfnamefont {L.}~\bibnamefont {Galgani}},\ and\ \bibinfo {author} {\bibfnamefont {A.}~\bibnamefont {Giorgilli}},\ }\bibfield  {title} {\bibinfo {title} {The {F}ermi-{P}asta-{U}lam problem and the metastability perspective},\ }\href@noop {} {\bibfield  {journal} {\bibinfo  {journal} {The Fermi-Pasta-Ulam problem: a status report}\ ,\ \bibinfo {pages} {151}} (\bibinfo {year} {2008})}\BibitemShut {NoStop}%
\bibitem [{\citenamefont {Chirikov}\ and\ \citenamefont {Izrailev}(1970)}]{chirikov1970statistical}%
  \BibitemOpen
  \bibfield  {author} {\bibinfo {author} {\bibfnamefont {B.~V.}\ \bibnamefont {Chirikov}}\ and\ \bibinfo {author} {\bibfnamefont {F.~M.}\ \bibnamefont {Izrailev}},\ }\bibfield  {title} {\bibinfo {title} {Statistical properties of a nonlinear string},\ }\href@noop {} {\bibfield  {journal} {\bibinfo  {journal} {Soviet Physics Doklady}\ }\textbf {\bibinfo {volume} {11}},\ \bibinfo {pages} {30} (\bibinfo {year} {1970})}\BibitemShut {NoStop}%
\bibitem [{\citenamefont {Nishida}(1971)}]{nishida1971note}%
  \BibitemOpen
  \bibfield  {author} {\bibinfo {author} {\bibfnamefont {T.}~\bibnamefont {Nishida}},\ }\bibfield  {title} {\bibinfo {title} {A note on an existence of conditionally periodic oscillation in a one-dimensional anharmonic lattice},\ }\href@noop {} {\bibfield  {journal} {\bibinfo  {journal} {Memoirs of the Faculty of Engineering, Kyoto University}\ }\textbf {\bibinfo {volume} {33}},\ \bibinfo {pages} {27} (\bibinfo {year} {1971})}\BibitemShut {NoStop}%
\bibitem [{\citenamefont {Rink}\ and\ \citenamefont {Verhulst}(2000)}]{rink2000near}%
  \BibitemOpen
  \bibfield  {author} {\bibinfo {author} {\bibfnamefont {B.}~\bibnamefont {Rink}}\ and\ \bibinfo {author} {\bibfnamefont {F.}~\bibnamefont {Verhulst}},\ }\bibfield  {title} {\bibinfo {title} {Near-integrability of periodic {FPU}-chains},\ }\href@noop {} {\bibfield  {journal} {\bibinfo  {journal} {Physica A: Statistical Mechanics and its Applications}\ }\textbf {\bibinfo {volume} {285}},\ \bibinfo {pages} {467} (\bibinfo {year} {2000})}\BibitemShut {NoStop}%
\bibitem [{\citenamefont {Rink}(2001)}]{rink2001symmetry}%
  \BibitemOpen
  \bibfield  {author} {\bibinfo {author} {\bibfnamefont {B.}~\bibnamefont {Rink}},\ }\bibfield  {title} {\bibinfo {title} {Symmetry and resonance in periodic {FPU} chains},\ }\href@noop {} {\bibfield  {journal} {\bibinfo  {journal} {Communications in Mathematical Physics}\ }\textbf {\bibinfo {volume} {218}},\ \bibinfo {pages} {665} (\bibinfo {year} {2001})}\BibitemShut {NoStop}%
\bibitem [{\citenamefont {Verhulst}(2020)}]{verhulst2020variations}%
  \BibitemOpen
  \bibfield  {author} {\bibinfo {author} {\bibfnamefont {F.}~\bibnamefont {Verhulst}},\ }\bibfield  {title} {\bibinfo {title} {Variations on the {Fermi-Pasta-Ulam} chain, a survey},\ }in\ \href@noop {} {\emph {\bibinfo {booktitle} {Chaotic Modeling and Simulation International Conference}}}\ (\bibinfo {organization} {Springer},\ \bibinfo {year} {2020})\ pp.\ \bibinfo {pages} {1025--1042}\BibitemShut {NoStop}%
\bibitem [{\citenamefont {Casetti}\ \emph {et~al.}(1997)\citenamefont {Casetti}, \citenamefont {Cerruti-Sola}, \citenamefont {Pettini},\ and\ \citenamefont {Cohen}}]{casetti1997fermi}%
  \BibitemOpen
  \bibfield  {author} {\bibinfo {author} {\bibfnamefont {L.}~\bibnamefont {Casetti}}, \bibinfo {author} {\bibfnamefont {M.}~\bibnamefont {Cerruti-Sola}}, \bibinfo {author} {\bibfnamefont {M.}~\bibnamefont {Pettini}},\ and\ \bibinfo {author} {\bibfnamefont {E.}~\bibnamefont {Cohen}},\ }\bibfield  {title} {\bibinfo {title} {{The Fermi-Pasta-Ulam problem revisited: Stochasticity thresholds in nonlinear Hamiltonian systems}},\ }\href@noop {} {\bibfield  {journal} {\bibinfo  {journal} {Physical Review E}\ }\textbf {\bibinfo {volume} {55}},\ \bibinfo {pages} {6566} (\bibinfo {year} {1997})}\BibitemShut {NoStop}%
\bibitem [{\citenamefont {Flach}\ \emph {et~al.}(2005)\citenamefont {Flach}, \citenamefont {Ivanchenko},\ and\ \citenamefont {Kanakov}}]{flach2005q}%
  \BibitemOpen
  \bibfield  {author} {\bibinfo {author} {\bibfnamefont {S.}~\bibnamefont {Flach}}, \bibinfo {author} {\bibfnamefont {M.}~\bibnamefont {Ivanchenko}},\ and\ \bibinfo {author} {\bibfnamefont {O.}~\bibnamefont {Kanakov}},\ }\bibfield  {title} {\bibinfo {title} {{q-Breathers and the Fermi-Pasta-Ulam problem}},\ }\href@noop {} {\bibfield  {journal} {\bibinfo  {journal} {Physical review letters}\ }\textbf {\bibinfo {volume} {95}},\ \bibinfo {pages} {064102} (\bibinfo {year} {2005})}\BibitemShut {NoStop}%
\bibitem [{\citenamefont {Flach}\ \emph {et~al.}(2006)\citenamefont {Flach}, \citenamefont {Ivanchenko},\ and\ \citenamefont {Kanakov}}]{flach2006q}%
  \BibitemOpen
  \bibfield  {author} {\bibinfo {author} {\bibfnamefont {S.}~\bibnamefont {Flach}}, \bibinfo {author} {\bibfnamefont {M.}~\bibnamefont {Ivanchenko}},\ and\ \bibinfo {author} {\bibfnamefont {O.}~\bibnamefont {Kanakov}},\ }\bibfield  {title} {\bibinfo {title} {{Q-breathers in Fermi-Pasta-Ulam chains: Existence, localization, and stability}},\ }\href@noop {} {\bibfield  {journal} {\bibinfo  {journal} {Physical Review E—Statistical, Nonlinear, and Soft Matter Physics}\ }\textbf {\bibinfo {volume} {73}},\ \bibinfo {pages} {036618} (\bibinfo {year} {2006})}\BibitemShut {NoStop}%
\bibitem [{\citenamefont {Chechin}\ \emph {et~al.}(2002)\citenamefont {Chechin}, \citenamefont {Novikova},\ and\ \citenamefont {Abramenko}}]{chechin2002bushes}%
  \BibitemOpen
  \bibfield  {author} {\bibinfo {author} {\bibfnamefont {G.}~\bibnamefont {Chechin}}, \bibinfo {author} {\bibfnamefont {N.}~\bibnamefont {Novikova}},\ and\ \bibinfo {author} {\bibfnamefont {A.}~\bibnamefont {Abramenko}},\ }\bibfield  {title} {\bibinfo {title} {{Bushes of vibrational modes for Fermi-Pasta-Ulam chains}},\ }\href@noop {} {\bibfield  {journal} {\bibinfo  {journal} {Physica D: Nonlinear Phenomena}\ }\textbf {\bibinfo {volume} {166}},\ \bibinfo {pages} {208} (\bibinfo {year} {2002})}\BibitemShut {NoStop}%
\bibitem [{\citenamefont {Eichhorn}\ \emph {et~al.}(2001)\citenamefont {Eichhorn}, \citenamefont {Linz},\ and\ \citenamefont {H{\"a}nggi}}]{eichhorn2001transformation}%
  \BibitemOpen
  \bibfield  {author} {\bibinfo {author} {\bibfnamefont {R.}~\bibnamefont {Eichhorn}}, \bibinfo {author} {\bibfnamefont {S.~J.}\ \bibnamefont {Linz}},\ and\ \bibinfo {author} {\bibfnamefont {P.}~\bibnamefont {H{\"a}nggi}},\ }\bibfield  {title} {\bibinfo {title} {Transformation invariance of {L}yapunov exponents},\ }\href@noop {} {\bibfield  {journal} {\bibinfo  {journal} {Chaos, Solitons \& Fractals}\ }\textbf {\bibinfo {volume} {12}},\ \bibinfo {pages} {1377} (\bibinfo {year} {2001})}\BibitemShut {NoStop}%
\bibitem [{\citenamefont {Flach}\ and\ \citenamefont {Ponno}(2008)}]{flach2008fermi}%
  \BibitemOpen
  \bibfield  {author} {\bibinfo {author} {\bibfnamefont {S.}~\bibnamefont {Flach}}\ and\ \bibinfo {author} {\bibfnamefont {A.}~\bibnamefont {Ponno}},\ }\bibfield  {title} {\bibinfo {title} {{The Fermi-Pasta-Ulam problem: Periodic orbits, normal forms and resonance overlap criteria}},\ }\href@noop {} {\bibfield  {journal} {\bibinfo  {journal} {Physica D: Nonlinear Phenomena}\ }\textbf {\bibinfo {volume} {237}},\ \bibinfo {pages} {908} (\bibinfo {year} {2008})}\BibitemShut {NoStop}%
\bibitem [{\citenamefont {Cretegny}\ \emph {et~al.}(1998)\citenamefont {Cretegny}, \citenamefont {Dauxois}, \citenamefont {Ruffo},\ and\ \citenamefont {Torcini}}]{cretegny1998localization}%
  \BibitemOpen
  \bibfield  {author} {\bibinfo {author} {\bibfnamefont {T.}~\bibnamefont {Cretegny}}, \bibinfo {author} {\bibfnamefont {T.}~\bibnamefont {Dauxois}}, \bibinfo {author} {\bibfnamefont {S.}~\bibnamefont {Ruffo}},\ and\ \bibinfo {author} {\bibfnamefont {A.}~\bibnamefont {Torcini}},\ }\bibfield  {title} {\bibinfo {title} {Localization and equipartition of energy in the $\beta$-{FPU} chain: {C}haotic breathers},\ }\href@noop {} {\bibfield  {journal} {\bibinfo  {journal} {Physica D: Nonlinear Phenomena}\ }\textbf {\bibinfo {volume} {121}},\ \bibinfo {pages} {109} (\bibinfo {year} {1998})}\BibitemShut {NoStop}%
\bibitem [{\citenamefont {Danieli}\ \emph {et~al.}(2017)\citenamefont {Danieli}, \citenamefont {Campbell},\ and\ \citenamefont {Flach}}]{danieli2017intermittent}%
  \BibitemOpen
  \bibfield  {author} {\bibinfo {author} {\bibfnamefont {C.}~\bibnamefont {Danieli}}, \bibinfo {author} {\bibfnamefont {D.}~\bibnamefont {Campbell}},\ and\ \bibinfo {author} {\bibfnamefont {S.}~\bibnamefont {Flach}},\ }\bibfield  {title} {\bibinfo {title} {Intermittent many-body dynamics at equilibrium},\ }\href@noop {} {\bibfield  {journal} {\bibinfo  {journal} {Physical Review E}\ }\textbf {\bibinfo {volume} {95}},\ \bibinfo {pages} {060202} (\bibinfo {year} {2017})}\BibitemShut {NoStop}%
\bibitem [{\citenamefont {Goedde}\ \emph {et~al.}(1992)\citenamefont {Goedde}, \citenamefont {Lichtenberg},\ and\ \citenamefont {Lieberman}}]{goedde1992chaos}%
  \BibitemOpen
  \bibfield  {author} {\bibinfo {author} {\bibfnamefont {C.~G.}\ \bibnamefont {Goedde}}, \bibinfo {author} {\bibfnamefont {A.~J.}\ \bibnamefont {Lichtenberg}},\ and\ \bibinfo {author} {\bibfnamefont {M.~A.}\ \bibnamefont {Lieberman}},\ }\bibfield  {title} {\bibinfo {title} {Chaos and the approach to equilibrium in a discrete sine-{G}ordon equation},\ }\href@noop {} {\bibfield  {journal} {\bibinfo  {journal} {Physica D: Nonlinear Phenomena}\ }\textbf {\bibinfo {volume} {59}},\ \bibinfo {pages} {200} (\bibinfo {year} {1992})}\BibitemShut {NoStop}%
\bibitem [{\citenamefont {Oseledec}(1968)}]{oseledec1968multiplicative}%
  \BibitemOpen
  \bibfield  {author} {\bibinfo {author} {\bibfnamefont {V.~I.}\ \bibnamefont {Oseledec}},\ }\bibfield  {title} {\bibinfo {title} {A multiplicative ergodic theorem, {L}yapunov characteristic numbers for dynamical systems},\ }\href@noop {} {\bibfield  {journal} {\bibinfo  {journal} {Transactions of the Moscow Mathematical Society}\ }\textbf {\bibinfo {volume} {19}},\ \bibinfo {pages} {197} (\bibinfo {year} {1968})}\BibitemShut {NoStop}%
\bibitem [{\citenamefont {Pesin}(1977)}]{pesin1977characteristic}%
  \BibitemOpen
  \bibfield  {author} {\bibinfo {author} {\bibfnamefont {Y.~B.}\ \bibnamefont {Pesin}},\ }\bibfield  {title} {\bibinfo {title} {Characteristic {L}yapunov exponents and smooth ergodic theory},\ }\href@noop {} {\bibfield  {journal} {\bibinfo  {journal} {Russian Mathematical Surveys}\ }\textbf {\bibinfo {volume} {32}},\ \bibinfo {pages} {55} (\bibinfo {year} {1977})}\BibitemShut {NoStop}%
\bibitem [{\citenamefont {Benettin}\ \emph {et~al.}(1980)\citenamefont {Benettin}, \citenamefont {Galgani}, \citenamefont {Giorgilli},\ and\ \citenamefont {Strelcyn}}]{benettin1980lyapunov}%
  \BibitemOpen
  \bibfield  {author} {\bibinfo {author} {\bibfnamefont {G.}~\bibnamefont {Benettin}}, \bibinfo {author} {\bibfnamefont {L.}~\bibnamefont {Galgani}}, \bibinfo {author} {\bibfnamefont {A.}~\bibnamefont {Giorgilli}},\ and\ \bibinfo {author} {\bibfnamefont {J.-M.}\ \bibnamefont {Strelcyn}},\ }\bibfield  {title} {\bibinfo {title} {{Lyapunov characteristic exponents for smooth dynamical systems and for Hamiltonian systems; a method for computing all of them. Part 1: Theory}},\ }\href@noop {} {\bibfield  {journal} {\bibinfo  {journal} {Meccanica}\ }\textbf {\bibinfo {volume} {15}},\ \bibinfo {pages} {9} (\bibinfo {year} {1980})}\BibitemShut {NoStop}%
\bibitem [{\citenamefont {Geist}\ \emph {et~al.}(1990)\citenamefont {Geist}, \citenamefont {Parlitz},\ and\ \citenamefont {Lauterborn}}]{geist1990comparison}%
  \BibitemOpen
  \bibfield  {author} {\bibinfo {author} {\bibfnamefont {K.}~\bibnamefont {Geist}}, \bibinfo {author} {\bibfnamefont {U.}~\bibnamefont {Parlitz}},\ and\ \bibinfo {author} {\bibfnamefont {W.}~\bibnamefont {Lauterborn}},\ }\bibfield  {title} {\bibinfo {title} {Comparison of different methods for computing {L}yapunov exponents},\ }\href@noop {} {\bibfield  {journal} {\bibinfo  {journal} {Progress of theoretical physics}\ }\textbf {\bibinfo {volume} {83}},\ \bibinfo {pages} {875} (\bibinfo {year} {1990})}\BibitemShut {NoStop}%
\bibitem [{\citenamefont {Abraham}\ and\ \citenamefont {Marsden}(2008)}]{abraham2008foundations}%
  \BibitemOpen
  \bibfield  {author} {\bibinfo {author} {\bibfnamefont {R.}~\bibnamefont {Abraham}}\ and\ \bibinfo {author} {\bibfnamefont {J.~E.}\ \bibnamefont {Marsden}},\ }\href@noop {} {\emph {\bibinfo {title} {Foundations of mechanics}}},\ \bibinfo {number} {364}\ (\bibinfo  {publisher} {American Mathematical Soc.},\ \bibinfo {year} {2008})\BibitemShut {NoStop}%
\bibitem [{\citenamefont {Palis}\ and\ \citenamefont {De~Melo}(2012)}]{palis2012geometric}%
  \BibitemOpen
  \bibfield  {author} {\bibinfo {author} {\bibfnamefont {J.~J.}\ \bibnamefont {Palis}}\ and\ \bibinfo {author} {\bibfnamefont {W.}~\bibnamefont {De~Melo}},\ }\href@noop {} {\emph {\bibinfo {title} {Geometric theory of dynamical systems: an introduction}}}\ (\bibinfo  {publisher} {Springer Science \& Business Media},\ \bibinfo {year} {2012})\BibitemShut {NoStop}%
\bibitem [{\citenamefont {Bivins}\ \emph {et~al.}(1973)\citenamefont {Bivins}, \citenamefont {Metropolis},\ and\ \citenamefont {Pasta}}]{bivins1973nonlinear}%
  \BibitemOpen
  \bibfield  {author} {\bibinfo {author} {\bibfnamefont {R.}~\bibnamefont {Bivins}}, \bibinfo {author} {\bibfnamefont {N.}~\bibnamefont {Metropolis}},\ and\ \bibinfo {author} {\bibfnamefont {J.~R.}\ \bibnamefont {Pasta}},\ }\bibfield  {title} {\bibinfo {title} {Nonlinear coupled oscillators: Modal equation approach},\ }\href@noop {} {\bibfield  {journal} {\bibinfo  {journal} {Journal of Computational Physics}\ }\textbf {\bibinfo {volume} {12}},\ \bibinfo {pages} {65} (\bibinfo {year} {1973})}\BibitemShut {NoStop}%
\bibitem [{\citenamefont {Sholl}(1990)}]{sholl1990modal}%
  \BibitemOpen
  \bibfield  {author} {\bibinfo {author} {\bibfnamefont {D.}~\bibnamefont {Sholl}},\ }\bibfield  {title} {\bibinfo {title} {Modal coupling in one-dimensional anharmonic lattices},\ }\href@noop {} {\bibfield  {journal} {\bibinfo  {journal} {Physics Letters A}\ }\textbf {\bibinfo {volume} {149}},\ \bibinfo {pages} {253} (\bibinfo {year} {1990})}\BibitemShut {NoStop}%
\bibitem [{\citenamefont {Sholl}\ and\ \citenamefont {Henry}(1991)}]{sholl1991recurrence}%
  \BibitemOpen
  \bibfield  {author} {\bibinfo {author} {\bibfnamefont {D.~S.}\ \bibnamefont {Sholl}}\ and\ \bibinfo {author} {\bibfnamefont {B.}~\bibnamefont {Henry}},\ }\bibfield  {title} {\bibinfo {title} {{Recurrence times in cubic and quartic Fermi-Pasta-Ulam chains: A shifted-frequency perturbation treatment}},\ }\href@noop {} {\bibfield  {journal} {\bibinfo  {journal} {Physical Review A}\ }\textbf {\bibinfo {volume} {44}},\ \bibinfo {pages} {6364} (\bibinfo {year} {1991})}\BibitemShut {NoStop}%
\bibitem [{\citenamefont {McLachlan}\ and\ \citenamefont {Atela}(1992)}]{mclachlan1992accuracy}%
  \BibitemOpen
  \bibfield  {author} {\bibinfo {author} {\bibfnamefont {R.~I.}\ \bibnamefont {McLachlan}}\ and\ \bibinfo {author} {\bibfnamefont {P.}~\bibnamefont {Atela}},\ }\bibfield  {title} {\bibinfo {title} {The accuracy of symplectic integrators},\ }\href@noop {} {\bibfield  {journal} {\bibinfo  {journal} {Nonlinearity}\ }\textbf {\bibinfo {volume} {5}},\ \bibinfo {pages} {541} (\bibinfo {year} {1992})}\BibitemShut {NoStop}%
\bibitem [{\citenamefont {Matsuyama}\ and\ \citenamefont {Konishi}(2015)}]{matsuyama2015multistage}%
  \BibitemOpen
  \bibfield  {author} {\bibinfo {author} {\bibfnamefont {H.~J.}\ \bibnamefont {Matsuyama}}\ and\ \bibinfo {author} {\bibfnamefont {T.}~\bibnamefont {Konishi}},\ }\bibfield  {title} {\bibinfo {title} {Multistage slow relaxation in a {H}amiltonian system: {The Fermi-Pasta-Ulam model}},\ }\href@noop {} {\bibfield  {journal} {\bibinfo  {journal} {Physical Review E}\ }\textbf {\bibinfo {volume} {92}},\ \bibinfo {pages} {022917} (\bibinfo {year} {2015})}\BibitemShut {NoStop}%
\bibitem [{\citenamefont {Eckmann}\ and\ \citenamefont {Ruelle}(1985)}]{eckmann1985ergodic}%
  \BibitemOpen
  \bibfield  {author} {\bibinfo {author} {\bibfnamefont {J.-P.}\ \bibnamefont {Eckmann}}\ and\ \bibinfo {author} {\bibfnamefont {D.}~\bibnamefont {Ruelle}},\ }\bibfield  {title} {\bibinfo {title} {Ergodic theory of chaos and strange attractors},\ }\href@noop {} {\bibfield  {journal} {\bibinfo  {journal} {Reviews of modern physics}\ }\textbf {\bibinfo {volume} {57}},\ \bibinfo {pages} {617} (\bibinfo {year} {1985})}\BibitemShut {NoStop}%
\bibitem [{\citenamefont {Ginelli}\ \emph {et~al.}(2007)\citenamefont {Ginelli}, \citenamefont {Poggi}, \citenamefont {Turchi}, \citenamefont {Chat{\'e}}, \citenamefont {Livi},\ and\ \citenamefont {Politi}}]{ginelli2007characterizing}%
  \BibitemOpen
  \bibfield  {author} {\bibinfo {author} {\bibfnamefont {F.}~\bibnamefont {Ginelli}}, \bibinfo {author} {\bibfnamefont {P.}~\bibnamefont {Poggi}}, \bibinfo {author} {\bibfnamefont {A.}~\bibnamefont {Turchi}}, \bibinfo {author} {\bibfnamefont {H.}~\bibnamefont {Chat{\'e}}}, \bibinfo {author} {\bibfnamefont {R.}~\bibnamefont {Livi}},\ and\ \bibinfo {author} {\bibfnamefont {A.}~\bibnamefont {Politi}},\ }\bibfield  {title} {\bibinfo {title} {Characterizing dynamics with covariant {L}yapunov vectors},\ }\href@noop {} {\bibfield  {journal} {\bibinfo  {journal} {Physical review letters}\ }\textbf {\bibinfo {volume} {99}},\ \bibinfo {pages} {130601} (\bibinfo {year} {2007})}\BibitemShut {NoStop}%
\bibitem [{\citenamefont {Baldovin}\ \emph {et~al.}(2021)\citenamefont {Baldovin}, \citenamefont {Vulpiani},\ and\ \citenamefont {Gradenigo}}]{baldovin2021statistical}%
  \BibitemOpen
  \bibfield  {author} {\bibinfo {author} {\bibfnamefont {M.}~\bibnamefont {Baldovin}}, \bibinfo {author} {\bibfnamefont {A.}~\bibnamefont {Vulpiani}},\ and\ \bibinfo {author} {\bibfnamefont {G.}~\bibnamefont {Gradenigo}},\ }\bibfield  {title} {\bibinfo {title} {Statistical mechanics of an integrable system},\ }\href@noop {} {\bibfield  {journal} {\bibinfo  {journal} {Journal of Statistical Physics}\ }\textbf {\bibinfo {volume} {183}},\ \bibinfo {pages} {41} (\bibinfo {year} {2021})}\BibitemShut {NoStop}%
\bibitem [{\citenamefont {Lando}\ and\ \citenamefont {Flach}(2023)}]{lando2023thermalization}%
  \BibitemOpen
  \bibfield  {author} {\bibinfo {author} {\bibfnamefont {G.~M.}\ \bibnamefont {Lando}}\ and\ \bibinfo {author} {\bibfnamefont {S.}~\bibnamefont {Flach}},\ }\bibfield  {title} {\bibinfo {title} {Thermalization slowing down in multidimensional {J}osephson junction networks},\ }\href@noop {} {\bibfield  {journal} {\bibinfo  {journal} {Physical Review E}\ }\textbf {\bibinfo {volume} {108}},\ \bibinfo {pages} {L062301} (\bibinfo {year} {2023})}\BibitemShut {NoStop}%
\end{thebibliography}%

\end{document}